\PassOptionsToPackage{unicode}{hyperref}
\PassOptionsToPackage{hyphens}{url}
\PassOptionsToPackage{dvipsnames,svgnames,x11names}{xcolor}
\documentclass[
  12pt]{article}

\usepackage{amsmath,amssymb,amsthm}
\usepackage{iftex}
\ifPDFTeX
  \usepackage[T1]{fontenc}
  \usepackage[utf8]{inputenc}
  \usepackage{textcomp} 
\else 
  \usepackage{unicode-math}
  \defaultfontfeatures{Scale=MatchLowercase}
  \defaultfontfeatures[\rmfamily]{Ligatures=TeX,Scale=1}
\fi
\usepackage{lmodern}
\ifPDFTeX\else  
\fi
\IfFileExists{upquote.sty}{\usepackage{upquote}}{}
\IfFileExists{microtype.sty}{
  \usepackage[]{microtype}
  \UseMicrotypeSet[protrusion]{basicmath} 
}{}
\makeatletter
\@ifundefined{KOMAClassName}{
  \IfFileExists{parskip.sty}{%
    \usepackage{parskip}
  }{
    \setlength{\parindent}{0pt}
    \setlength{\parskip}{6pt plus 2pt minus 1pt}}
}{
  \KOMAoptions{parskip=half}}
\makeatother
\usepackage{xcolor}
\makeatletter
\ifx\paragraph\undefined\else
  \let\oldparagraph\paragraph
  \renewcommand{\paragraph}{
    \@ifstar
      \xxxParagraphStar
      \xxxParagraphNoStar
  }
  \newcommand{\xxxParagraphStar}[1]{\oldparagraph*{#1}\mbox{}}
  \newcommand{\xxxParagraphNoStar}[1]{\oldparagraph{#1}\mbox{}}
\fi
\ifx\subparagraph\undefined\else
  \let\oldsubparagraph\subparagraph
  \renewcommand{\subparagraph}{
    \@ifstar
      \xxxSubParagraphStar
      \xxxSubParagraphNoStar
  }
  \newcommand{\xxxSubParagraphStar}[1]{\oldsubparagraph*{#1}\mbox{}}
  \newcommand{\xxxSubParagraphNoStar}[1]{\oldsubparagraph{#1}\mbox{}}
\fi
\makeatother

\usepackage{longtable,booktabs,array}
\usepackage{calc} 
\usepackage{etoolbox}
\makeatletter
\patchcmd\longtable{\par}{\if@noskipsec\mbox{}\fi\par}{}{}
\makeatother
\IfFileExists{footnotehyper.sty}{\usepackage{footnotehyper}}{\usepackage{footnote}}
\makesavenoteenv{longtable}
\usepackage{graphicx}
\makeatletter
\def\maxwidth{\ifdim\Gin@nat@width>\linewidth\linewidth\else\Gin@nat@width\fi}
\def\maxheight{\ifdim\Gin@nat@height>\textheight\textheight\else\Gin@nat@height\fi}
\makeatother
\setkeys{Gin}{width=\maxwidth,height=\maxheight,keepaspectratio}
\makeatletter
\def\fps@figure{htbp}
\makeatother

\makeatletter
\@ifpackageloaded{caption}{}{\usepackage{caption}}
\AtBeginDocument{%
\ifdefined\contentsname
  \renewcommand*\contentsname{Table of contents}
\else
  \newcommand\contentsname{Table of contents}
\fi
\ifdefined\listfigurename
  \renewcommand*\listfigurename{List of Figures}
\else
  \newcommand\listfigurename{List of Figures}
\fi
\ifdefined\listtablename
  \renewcommand*\listtablename{List of Tables}
\else
  \newcommand\listtablename{List of Tables}
\fi
\ifdefined\figurename
  \renewcommand*\figurename{Figure}
\else
  \newcommand\figurename{Figure}
\fi
\ifdefined\tablename
  \renewcommand*\tablename{Table}
\else
  \newcommand\tablename{Table}
\fi
}
\@ifpackageloaded{float}{}{\usepackage{float}}
\floatstyle{ruled}
\@ifundefined{c@chapter}{\newfloat{codelisting}{h}{lop}}{\newfloat{codelisting}{h}{lop}[chapter]}
\floatname{codelisting}{Listing}

\makeatother
\makeatletter
\@ifpackageloaded{caption}{}{\usepackage{caption}}
\@ifpackageloaded{subcaption}{}{\usepackage{subcaption}}
\makeatother

\ifLuaTeX
  \usepackage{selnolig}  
\fi
\usepackage[]{natbib}
\usepackage{bookmark}

\IfFileExists{xurl.sty}{\usepackage{xurl}}{} 
\hypersetup{
  pdftitle={Fractional Ridge Regression},
  pdfauthor={Sihyung Park, Leonard A. Stefanski},
  pdfkeywords={3 to 6 keywords, that do not appear in the title},
  colorlinks=true,
  linkcolor={blue},
  filecolor={Maroon},
  citecolor={Blue},
  urlcolor={Blue},
  pdfcreator={LaTeX via pandoc}}

\newcommand{\anon}{1}

\usepackage{algorithm}
\usepackage{algpseudocode}
\usepackage{setspace}

\newtheorem{lemma}{Lemma}
\newtheorem{theorem}{Theorem}
\newtheorem{corollary}{Corollary}

\theoremstyle{remark}
\newtheorem*{remark}{Remark}

\newcommand{\tred}{}
 
\newcommand{\E}{\mathbb{E}}
\newcommand{\pr}{\mathbb{P}}

\newcommand{\wh}{\widehat}

\DeclareMathOperator{\sign}{sign}
\DeclareMathOperator{\sic}{SIC}

\newcommand{\bbeta}{ \boldsymbol{\beta}}

\newcommand{\bdelta}{ \boldsymbol{\delta}}

\newcommand{\bC}{ \mbox{\bf C}}

\newcommand{\bd}{ \mbox{\bf d}}

\newcommand{\bx}{ \mbox{\bf x}}
\newcommand{\bX}{\mathbf{X}}

\newcommand{\by}{ \mbox{\bf y}}

\newcommand{\bb}{ \mbox{\bf b}}

\newcommand{\bu}{ \mbox{\bf u}}

\newcommand{\bv}{ \mbox{\bf v}}
\newcommand{\bV}{ \mbox{\bf V}}

\newcommand{\bg}{ \mbox{\bf g}}
\newcommand{\bH}{ \mbox{\bf H}}
\newcommand{\bI}{ \mbox{\bf I}}

\newcommand{\bW}{ \mbox{\bf W}}
\newcommand{\bw}{ \mbox{\bf w}}

\newcommand{\bzero}{ \mbox{\bf 0}}

\newcommand{\argmin}{{\mathop{\rm arg\, min}}}

\newcommand{\beq}{ \begin{equation}}
\newcommand{\eeq}{ \end{equation}}
\newcommand{\beqn}{ \begin{eqnarray}}
\newcommand{\eeqn}{ \end{eqnarray}}

\usepackage{caption}
\begin{document}

\def\spacingset#1{\renewcommand{\baselinestretch}%
{#1}\small\normalsize} \spacingset{1}


\if1\anon
{
  \title{\bf Fractional Ridge Regression}
  \author{
    Sihyung Park\thanks{
    The authors gratefully acknowledge National Science Foundation for their funding of this research (Award Number: 2310208).}\hspace{.2cm}\\
    Department of Statistics, North Carolina State University\\
    and \\
    Leonard A. Stefanski$^*$ \\
    Department of Statistics, North Carolina State University}
  \maketitle
} \fi

\if0\anon
{
  \bigskip
  \bigskip
  \bigskip
  \begin{center}
    {\LARGE\bf Fractional Ridge Regression}
\end{center}
  \medskip
} \fi

\bigskip
\begin{abstract}
 Regularization methods such as the Lasso combine shrinkage and variable selection, but strong penalization drives their solutions toward the null model. We introduce Fractional Ridge (Fridge) regression, a penalized regression method that instead shrinks toward models of a prespecified target size. Its penalty vanishes whenever the number of nonzero coefficients does not exceed that target. 
This behavior connects continuous shrinkage with best subset selection. We develop an efficient recursive penalty-evaluation algorithm, coordinate descent updates, and a predictor-corrector algorithm for tracing solution paths. 
Under suitable conditions, we establish asymptotic normality, the oracle property, and convergence of global minimizers to best subset solutions under extreme regularization. Simulation studies demonstrate competitive prediction and variable selection across a range of signal strengths and predictor correlations. 
An application to riboflavin production data illustrates how Fridge selects a small set of predictors whose leading coefficients agree with signals prioritized by a Bayesian method developed specifically for genetic fine-mapping.
\end{abstract}

\noindent%
{\it Keywords:} Adaptive penalization; Best subset selection; Nonconvex optimization; Oracle property; Solution paths; Variable selection. 
\vfill

\newpage
\spacingset{1.8} 

\section{Introduction}
Regression with many predictors requires balancing the information retained in a model against the variability of its parameter estimates. Variable selection addresses this problem by excluding predictors, while shrinkage reduces the magnitude of their coefficients. Best subset selection, ridge regression, and the Lasso take different positions along this trade-off, motivating methods that combine sparse modeling with selective shrinkage \tred{\citep[Chapter 3]{hastie2009elements}}.

For a response vector $\by\in\mathbb R^n$ and design matrix $\bX\in\mathbb R^{n\times p}$, best subset selection solves
\begin{equation}
\hat\bbeta^{\mathrm{BSS}}\in\argmin_{\bbeta}\|\by-\bX\bbeta\|_2^2
\quad\text{subject to}\quad\|\bbeta\|_0\le k,
\label{eq:bss}
\end{equation}
where $\|\bbeta\|_0$ counts nonzero coefficients. The selected coefficients are fitted by least squares without further shrinkage, but small changes in the data can change the selected subset and increase prediction variability \citep{breiman1996heuristics}. Recent advances have expanded the range of problems for which subset selection is computationally practical \citep{bertsimas2016best,zhu2020polynomial}; nevertheless, the statistical trade-off between sparsity and stability remains.

Ridge regression addresses instability by adding the penalty $\lambda\sum_j\beta_j^2$ to the residual sum of squares \citep{hoerl1970ridge}. It stabilizes estimation with correlated predictors and gives a unique solution for $\lambda>0$, including when $p>n$, but generally retains all predictors. The Lasso uses $\lambda\sum_j|\beta_j|$ to combine shrinkage and variable selection in a convex optimization problem \citep{tibshirani1996regression}. Related approaches include bridge regression, with penalties $\sum_j|\beta_j|^q$ \citep{frank1993statistical}, and the nonnegative garrote, which shrinks preliminary coefficient estimates through constrained scaling factors \citep{breiman1995better}. Together, these methods provide ways to move between sparse selection and continuous shrinkage.

The Lasso's shrinkage introduces bias in large coefficients \tred{\citep{fan2001variable}}. Nonconvex penalties such as SCAD \citep{fan2001variable} and MCP \citep{zhang2010nearly} reduce this bias by allowing the penalty derivative to vanish for sufficiently large coefficients. Under suitable conditions, these estimators attain the oracle property: consistent variable selection and the same limiting distribution for the active coefficients as if the true model were known \tred{\citep{fan2001variable,zhang2010nearly}}. The adaptive Lasso pursues a similar goal through data-dependent weights, penalizing coefficients with larger preliminary estimates less heavily \citep{zou2006adaptive}. These approaches make penalization selective, either through the coefficient magnitude itself or through evidence from an initial fit.

This selectivity raises a question about the model toward which shrinkage is directed. With the usual fixed positive weights, ridge and Lasso penalties vanish only at the zero vector. Their solutions therefore approach the null model as regularization increases. SCAD and MCP protect large coefficients at a given tuning parameter, but do not prescribe a non-null target model under extreme regularization. A penalty that instead vanishes on all models of a specified maximum size would provide a different connection between shrinkage and subset selection.

\begin{figure}[htp]
  \includegraphics[width=\textwidth]{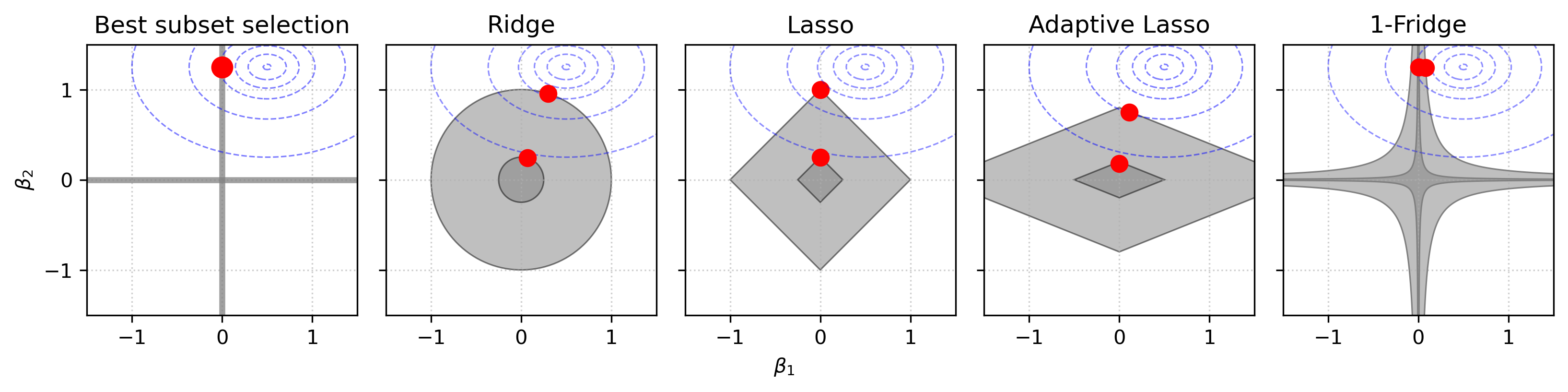}
  \caption[Geometry of select regression penalties ($p=2$).]{Regions of select regression penalties ($p=2$). Dashed lines show the OLS contour centered at $(0.5, 1.25)$. Shaded lines and shapes show penalty regions, with darker shades corresponding to stronger regularization. Dots show penalized solutions. For best subset selection, $k$ in \eqref{eq:bss} is set to one. Adaptive Lasso weights are set to $(1/0.5, 1/1.25)$, based on the OLS solution. \quad Best subset selection region covers only the axes, displaying its hard-thresholding property.  Ridge, Lasso, and adaptive Lasso solutions collapse to the zero vector as penalization gets stronger. Fractional Ridge with target model size 1 (1-Fridge), which is introduced in Section \ref{sec:def}, retains one or more non-zero coefficients even under extreme regularization.}
  \label{fig:penalties}
\end{figure}

\noindent \textbf{Contribution.} \quad
We introduce \textit{Fractional Ridge (Fridge)} regression, whose penalty vanishes whenever at most $m$ coefficients are nonzero. The parameter $m$ specifies a target model size, while $\lambda$ controls the strength of regularization toward that target. A fit at finite $\lambda$ may retain more than $m$ predictors, allowing shrinkage to operate without imposing a hard size constraint. Under the conditions established below, global minimizers approach best subset solutions as $\lambda\to\infty$. Figure \ref{fig:penalties} illustrates the geometric distinction for $p=2$: the Fridge penalty region retains coordinate subspaces as regularization increases.

We develop algorithms for evaluating the penalty, fitting Fridge models, and computing their solution paths. Our theoretical analysis establishes limiting distributions, oracle properties under the stated tuning conditions, and the connection to best subset selection. Simulations examine how these properties translate into prediction and selection across signal-to-noise ratios. A riboflavin application illustrates the behavior of Fridge with highly-correlated, gene-expression predictors.

The rest of the paper is organized as follows: Section \ref{sec:def} defines the penalty, Section \ref{sec:algo} develops the algorithms, and Section \ref{sec:theory} presents the theoretical results. Section \ref{sec:sim} reports simulations, followed by the riboflavin application (Section \ref{sec:ribo}) and a discussion in Section \ref{sec:discussion}.

\section{Fractional Ridge Penalty}  \label{sec:def}
\subsection{Definition}
The motivation for the Fridge penalty is most clearly described in low-dimensional settings. Consider a response variable $Y_i$ and $p=3$ predictors $\bX_i = (X_{i1}, X_{i2}, X_{i3})$ with corresponding coefficients $\bbeta = (\beta_1, \beta_2, \beta_3)^\top$. Assume that $(\bX_i, y_i)$, $i=1,\dots,n$ are independent and identically distributed, and that, without loss of generality, both the response and predictors are centered.

Define the sum of squared errors as $Q(\bbeta) = {n}^{-1} \|\by - \bX\bbeta\|_2^2$ and the ridge penalty as $P_{0}(\bbeta) = \sum_{j=1}^3 \beta_j^2 = \|\bbeta\|_2^2$. The ridge estimator $\wh\bbeta_0(\lambda)$ is then given by
$$
\wh{\bbeta}_0(\lambda) = \argmin_{\bbeta} \left\{ Q(\bbeta) + \lambda P_{0}(\bbeta) \right \}.
$$
Under extreme regularization, the ridge estimator shrinks to the zero vector: $\lim_{\lambda \rightarrow \infty} \wh\bbeta_0(\lambda) = \bzero_{3\times 1}$. This is because $P_0(\bbeta) = 0$ if and only if $\bbeta=0$.

Now, consider the alternative estimator $\wh\bbeta_1(\lambda)$ defined by
\begin{align}
\wh{\bbeta}_1(\lambda) = \argmin_{\bbeta} \left\{ Q(\bbeta) + \lambda P_{1}(\bbeta) \right \}, \label{fridge1}
\end{align}
where $P_{1}(\bbeta) = \beta_1^2\beta_2^2 + \beta_1^2\beta_3^2 + \beta_2^2\beta_3^2$.   Observe that $P_{1}(\bbeta) = 0$ if and only if at least two of the $\beta_j$ are zero; notably, it does not require all $\beta_j$ to be zero. The solution $\wh\bbeta_1(\lambda)$ that minimizes $Q(\bbeta)$ subject to $P_{1}(\bbeta) = 0$ corresponds to the best single-variable model in terms of the smallest sum of squared errors. Since $\lim_{\lambda \rightarrow \infty} \wh\bbeta_1(\lambda)$ minimizes $Q(\bbeta)$ subject to $P_{1}(\bbeta) = 0$, the Fridge estimator $\wh\bbeta_1(\lambda)$, under extreme regularization, shrinks towards this best one-variable model, not towards the null model.
Similarly, one can consider the penalty $P_2(\bbeta) = \beta_1^2\beta_2^2\beta_3^2$ to shrink towards the best two-variable model. In this case, $P_2(\bbeta) = 0$ if and only if at least one $\beta_j$ is zero, allowing as many as two non-zero coefficient estimates.   

In general, to induce shrinkage towards the best subset of size $m < p$, the Fridge penalty is
\begin{align}
\label{eq:gen_pen_def}
P_{m}(\bg(\bbeta)) = \sum_{1 \leq j_1 < j_2 < \cdots < j_{m+1} \leq p} g_{j_1}(\beta_{j_1}) g_{j_2}(\beta_{j_2}) \cdots g_{j_{m+1}}(\beta_{j_{m+1}}),
~~\text{for}~~ 0 \leq m < p,
\end{align}
where $\bg(\bbeta) = (g_1(\beta_1), \dots, g_p(\beta_p))^\top$ and $g_j(\beta_{j})$ is such that $g_j(b) \ge 0$ for all $b$ and $g_j(b) = 0$ if and only if $b=0$, e.g., $g_j(b) = b^2$ or $|b|$.
By construction, $P_{m}(\bg(\bbeta))$ is zero if and only if at least $(p-m)$ of the $\beta_j$ are zero. For any $\lambda > 0$, the corresponding Fridge estimator $\wh{\bbeta}_m(\lambda)$ is defined as the solution to $\argmin_{\bbeta} \{ Q(\bbeta) + \lambda P_{m}(\bg(\bbeta)) \}$. As $\lambda \to \infty$, $\wh{\bbeta}_m(\lambda)$ shrinks towards the best $m$-variable model.

$\wh\bbeta_m(\lambda)$ is termed an $m$-\textit{Fractional Ridge} estimator because it is designed to retain a fraction ($m/p$) of coefficients as potentially non-zero, rather than shrinking all coefficients towards zero. We will refer to $m$ as the \textit{target model size (TMS)}, as $\wh\bbeta_m(\lambda)$ shrinks towards a model of at most size $m$ under extreme regularization. With this definition, ridge, Lasso, and other methods that shrink coefficients towards zero (i.e., towards a null model) have a TMS of $0$.

The choice $g_j(b) = b^2$ arises naturally as a direct generalization of the ridge regression. For the algorithms described in later sections and for better finite-sample properties, we use $g_j(b) = |b|$ or $g_j(b) = |b| / |\wh\beta_j^\text{init}|$ for some initial estimate $\wh\bbeta^\text{init}$. In this sense, Fridge is complementary rather than directly competitive to existing methods, as $g_j(\cdot)$ can be any existing penalty function, e.g., SCAD, MCP. For the remaining sections, we fix $g_j(b) = |b|/|\wh\beta_j^\text{init}|$ and $P_m(\bbeta) := P_m(\bg(\bbeta))$ with slight abuse of notation unless $g_j$ is otherwise stated. 

Our primary interest lies not in achieving the target model size $m$ itself, but rather in identifying models along the solution path $\{\wh\bbeta_m(\lambda) : \lambda > 0\}$ that exhibit optimal prediction-generalizability balance.

\subsection{Geometry}
As touched upon in its definition, a key distinction between Fridge and other Lasso-type penalties lies in their behavior under extreme regularization. To illustrate these characteristics, consider the constrained optimization problem:
\begin{align}
	\min_{\bbeta} ~ Q(\bbeta) \quad \textrm{subject to} ~ P_m(\bbeta) \leq \tau.
\end{align}
For easier visualization, we assume $\wh\bbeta^\text{init}$ to be a vector of ones, and consider the $1$-Fridge penalty $P_1(\bbeta)$.

For $p=2$ case with $\bbeta = (\beta_1, \beta_2)^\top$, the penalty regions depicted in Figure \ref{fig:penalties} demonstrate the distinct behavior of Fridge as the constraint parameter $\tau$ approaches zero (or equivalently, $\lambda\to\infty$). Unlike the Lasso, which forces all coefficients towards zero, Fridge can selectively shrink a fraction of the coefficient estimates. This is evidenced by the small change in $\beta_2$ values along Fridge's penalty contour, even when $\beta_1$ is heavily penalized (Figure \ref{fig:penalties}). In the most extreme regularization case where $\tau = 0$, the Lasso estimate is invariably $(0,0)$, regardless of the underlying data. In contrast, the Fridge estimate can still possess a non-zero component. Specifically, the Fridge penalty region always includes the axes. This unique property allows Fridge to preserve truly active features even under strong regularization.

\begin{figure}[!htp]
\centering
  \includegraphics[width=0.21\textwidth]{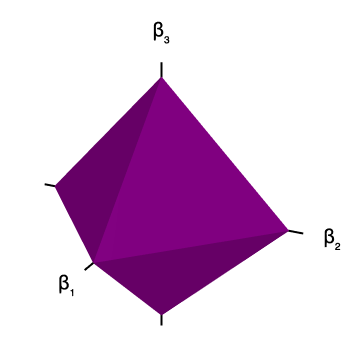}
  \includegraphics[width=0.24\textwidth]{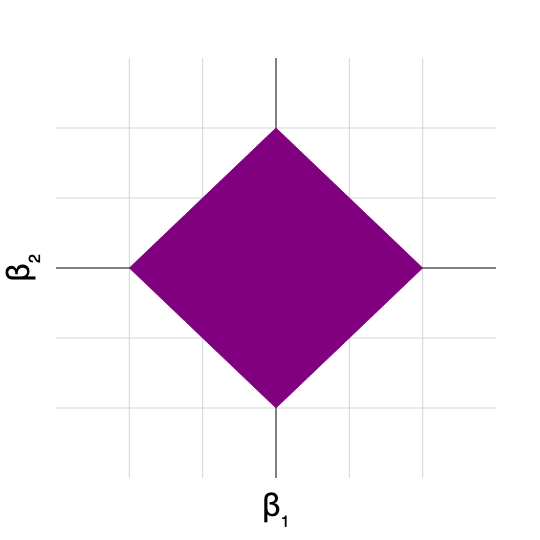}
  \includegraphics[width=0.24\textwidth]{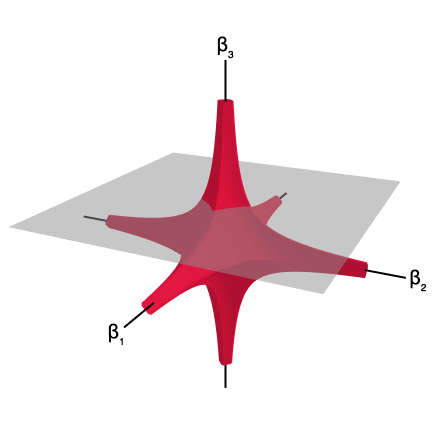}
  \includegraphics[width=0.24\textwidth]{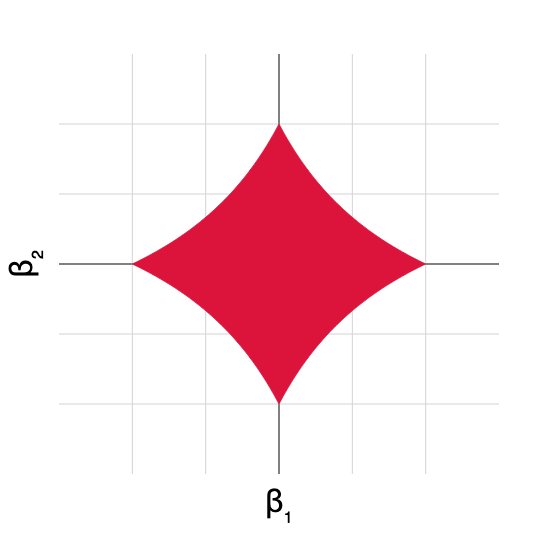}
  
  (a) \qquad\qquad\qquad\quad\;\;\; 
  (b) \qquad\qquad\quad\qquad\;\;\; 
  (c) \qquad\qquad\quad\qquad\;\;\; 
  (d) \quad
  
  \caption[Geometry of the Lasso and 1-Fridge penalties with three covariates.]{
    Geometry of the Lasso and 1-Fridge penalties with three covariates. Plot (a) illustrates the Lasso penalty region; (b) displays the cross-section of the Lasso region after fixing $\beta_3$ to a small value; (c) illustrates 1-Fridge penalty region; (d) shows the cross-section of (c) after fixing $\beta_3$ to a non-zero value.
  }
  \label{fig:penalty3d}
\end{figure}

For $p=3$ case with $\bbeta = (\beta_1, \beta_2, \beta_3)^\top$ (Figure \ref{fig:penalty3d}), if $\beta_3$ is fixed at one, the penalty constraint on remaining $\beta_1$ and $\beta_2$ becomes $\lambda(|\beta_1|+|\beta_2|+|\beta_1\beta_2|) \leq \tau$. This corresponds to a star shape on the $\beta_1 \times \beta_2$ plane, which now exhibits ``corners,'' similar to how the Lasso penalty behaves (Figure \ref{fig:penalty3d}(d)).    Generally, for $g_j(b) = |b|/|\wh\beta_j^\text{init}|$, if $m$ coefficients are held constant at non-zero values, the region defined by $P_m(\bbeta) \leq \tau$ for the remaining $(p-m)$ coefficients no longer includes the coordinate axes. Instead, this constrained region displays sharp corners where the optimization path naturally leads to a zero estimate for one of the remaining coefficients.   This property suggests that a coordinate descent algorithm applied to Fridge will yield sparse solutions. 

Finally, the Fridge penalty 
encourages sparse solution when predictors are correlated. For an illustrative example, consider a data generating process $\mathbb E[y|\bX]=X_1+X_2$ where $X_1$ and $X_2$ are perfectly correlated and on the same scale. The Lasso penalty, represented as $|\hat\beta_1|+|\hat\beta_2|$, yields infinitely many solutions; ridge penalty $\hat\beta_1^2+\hat\beta_2^2$ prefers dense solution with equal magnitudes. In contrast, the $1$-Fridge penalty $P_1(\bbeta) = |\beta_1\beta_2|$ achieves its minimum when one of the coefficients is zero.

\subsection{Fractional Ridge as a Weighted Lasso Problem}  \label{sec:weighted}

The adaptive Lasso utilizes weights derived from an initial estimator to provide a more consistent variable selection process. These weights are typically defined as $w_j = {1}/{|\hat{\beta}_j^{\text{init}}|^{\gamma}}$ for positive $\gamma$ \tred{\citep{zou2006adaptive}}. 
By this formulation, the model ensures that predictors exhibiting a small initial signal (low $|\hat{\beta}_j^{\text{init}}|$) are assigned larger weights, thereby increasing the probability of their coefficients being penalized to zero. Conversely, predictors with strong initial evidence are preserved with minimal bias. Notably, the adaptive Lasso weight $w_j^{\text{adaLasso}}$ can approach infinity if the initial estimate is near zero, effectively forcing the exclusion of that variable from the final model.

In contrast, the {Fridge} penalty introduces a collaborative, or ``all-against-all,'' competition among predictors. Under this regime, the Fridge penalty can be viewed as an adaptively weighted Lasso with weights proportional to the magnitudes of other coefficients in the model, creating an interdependent penalty structure:
\begin{equation}
w_j^{\text{fridge}} \propto \sum_{\substack{1 \leq j_1 < j_2 < \dots < j_m \leq p \\ j_1, \dots, j_m \neq j}} \frac{|\beta_{j_1} \dots \beta_{j_m}|}{|\wh\beta_{j_1}^\text{init} \dots \wh\beta_{j_m}^\text{init}|} \frac1{|\hat\beta_j^\text{init}|}.
\end{equation}
This approach additionally punishes weak predictors based on their relative magnitude compared to the broader feature set. 
A critical distinction between these two methods lies in the behavior of their weights at the limit. While the adaptive Lasso weight can reach infinity to exclude a variable, the Fridge weight $w_j^{\text{fridge}}$ can be zero, leaving a variable weakly or not penalized.

\section{Algorithm}  \label{sec:algo}
There are two challenges in evaluating the Fridge penalty and solving the penalized optimization problem. 
First, the penalty $P_m(\bg(\bbeta))$ involves $p$-choose-$(m+1)$ terms, which grows at least as fast as $O(p^m)$. Without a proper approach, even evaluating a single penalty term would cost heavy computation. 
Second, the Fridge objective function is nonconvex. This necessitates an algorithm for efficient computation and conditions under which the algorithm guarantees convergence to the solution.
\subsection{Penalty Evaluation}
We demonstrate that the $m$-Fridge penalty $P_m(\bg(\bbeta))$ can be computed with $\text{O}( mp )$ arithmetic operations via a \textit{forward recursive algorithm}. This makes Fridge feasible for values of $m$ and $p$ where enumerating all $\binom{p}{m+1}$ possible combinations of products of $(m+1)$ coefficients would be computationally infeasible.

For $\bg(\bbeta) \in \mathbb{R}^p$,  define $\bV_{0}(\bg(\bbeta))=\bg(\bbeta)$. It is straightforward to see that for $m=0$, $P_{0}(\bg(\bbeta)) = \mathbf{1}^\top \bg(\bbeta) = \mathbf{1}^\top \bV_{0}(\bg(\bbeta))$. For $k=1,\dots, m < p$, we define the recursive relations:
\begin{align}
\label{eq:forward}
{\bV}_{k}(\bg(\bbeta)) &= \{ P_{k-1}(\bg(\bbeta)) - {\bV}_{k-1}(\bg(\bbeta)) \} \circ \bg(\bbeta), \\
P_{k}(\bg(\bbeta)) &= \frac{\mathbf{1}^\top{\bV}_{k}(\bg(\bbeta))}{k+1}.  \label{eq:tildePdk}
\end{align}
Here, $\circ$ denotes the Hadamard (element-wise) product.

Essentially, for $k \ge 2$, $P_{k}(\bg(\bbeta))$ represents the sum of all $\binom{p}{k+1}$ unique products of $(k+1)$ terms from the elements of $\bg$. The $j$-th element of $\bV_{k}(\bg(\bbeta))$ is $g_j(\beta_j)P_{k-1}(\bg(\bbeta)_{-[j]})$, which is $g_j(\beta_j)$ times the sum of all $\binom{p-1}{k}$ products of $k$ terms from the remaining elements. The case $k=1$ provides a familiar and informative illustration:
\begin{align*}
\mathbf{1}^\top \bV_{1}(\bg(\bbeta)) &= (\mathbf{1}^\top\bg(\bbeta))P_{0}(\bg(\bbeta)) - \mathbf{1}^\top\{\bV_{0}(\bg(\bbeta))\odot\bg(\bbeta) \} \\
&= (\mathbf{1}^\top\bg(\bbeta))^2 - \mathbf{1}^\top(\bg(\bbeta) \odot\bg(\bbeta)) \\
&= (g_1+\cdots + g_p)^2 - (g_1^2 + \cdots + g_p^2) \\
&= 2\sum_{i < j} g_ig_j,
\end{align*}
Dividing by $2$ yields $P_{1}(\bg(\bbeta)) = \sum_{i<j} g_i(\beta_i) g_j(\beta_j)$, which is consistent with the definition for $m=1$.

When counting arithmetic operations for the penalty $P_m(\bg(\bbeta))$ in \eqref{eq:tildePdk}, but ignoring operations required to calculate $g(\cdot)$, computing the Fridge penalty with the target model size $m$ requires $(3m+1)p-1$ operations with our algorithm, thus demonstrating linearity in both $m$ and $p$.

\subsection{Fitting a Fractional Ridge Model} 

For convex $g(\cdot)$, the Fridge penalty is coordinate-wise convex. Specifically, the penalty $P_{m}(\bg(\bbeta))$ can be expressed as $g_j(\beta_j)$ multiplied by a function of the remaining coefficients $\bbeta_{-[j]}$. Here, $\bbeta_{-[j_1,\dots,j_k]}$ denotes the reduced-dimension coefficient vector excluding $\beta_{j_1}, \dots, \beta_{j_k}$. 
This structure suggests a coordinate descent algorithm for fitting a Fridge model. Lemma \ref{lemma:derivative} further details this by presenting the partial derivatives of the Fridge penalty.

\begin{lemma} \label{lemma:derivative}
Let $P_k(\bg) = \sum_{1 \leq j_1 < j_2 < \cdots < j_{k+1} \leq p} g_{j_1} g_{j_2} \cdots g_{j_{k+1}}$ for $\bg \in \mathbb{R}^p$. Let $P_{-1} \equiv 1$. For $k=1,\dots,p-1$ and $j=1,\dots,p$, the partial derivatives of $P_k(\bg)$ are given by
	\begin{align*}
 \frac{\partial}{\partial g_j} P_{k}(\bg)  
   =P_{k-1}(\bg_{-[j]}), 
   \qquad
 \frac{\partial^2}{\partial g_j\partial g_r} P_{k}(\bg) = 
 \begin{cases}
 	 0,                     &   j = r,  \\
     P_{k-2}(\bg_{-[j,r]}), &   j\ne r.
 \end{cases}
\end{align*}
\end{lemma}

Calculating $P_{k-1}(\bg_{[-j]})$ and $P_{k-2}(\bg_{[-j,r]})$ each requires $O(kp)$ arithmetic operations, respectively. For $m$, the naive full gradient calculation requires $O(mp^2)$ arithmetic operations, and the Hessian, $O(mp^3)$ arithmetic operations.

To derive the coordinate descent (CD) update rule, we rewrite the Fridge objective function for target model size $m$ as $R_{\lambda,m}(\bg(\bbeta)) = Q(\bbeta)/2 + \lambda P_{m}(\bg(\bbeta))$. 
The coordinate descent update rule is formally stated in Theorem \ref{thm:cd}.
\begin{theorem}  \label{thm:cd}
	Let $S(z,\gamma) = \text{sign}(z)(|z|-\gamma)_+$ be the soft thresholding operator. The coordinate descent update rule for Fridge estimator $\wh\bbeta$ with $g_j(b) = w_j |b|$, $w_j > 0$ is
\begin{align}
	\hat\beta_j^{(\textrm{new})} \leftarrow
	S\left\{ \frac1n\sum_{i=1}^n x_{ij}(y_i - \hat y_i^{(j)}),~ \lambda w_j P_{m-1}(\hat{\bg}_{-[j]}) \right\}  \bigg/ \left( \frac1n\sum_{i=1}^n x_{ij}^2 \right), \quad j=1,\dots,p,
	\label{cdupdate_raw}
\end{align}
where $\wh{\by}^{(j)} = \bX_{-[j]}\wh{\bbeta}_{-[j]}$ represents the fitted value excluding the predictor $\bx_j$ and $\hat{\bg}=\bg(\wh\bbeta)$.
\end{theorem}

\begin{corollary} \label{cor:cd}
Suppose $S\subset\{1,\dots,p\}$ has $|S|=m$, $X_S$ has full column rank, and let $\hat\bbeta_S^{OLS} = (\bX_S^\top \bX_S)^{-1}\bX_S^\top \by$. Assume every component of $\hat\bbeta_S^{OLS}$ is nonzero.
Define
\begin{align}  \label{eq:lamcondition}
{ \lambda_S = \max_{j\notin S} \frac{ |\bx_j^\top (\by-\bX_S\hat\bbeta_S^{OLS})| }{ n w_j P_{m-1}(\bg_{-[j]}) }. }
\end{align}
For any ${\lambda\ge\lambda_S,} $
the vector of 
$$ \tilde\beta_j= \begin{cases} \hat\beta_j^{OLS},&j\in S,\\ 0,&j\notin S \end{cases} $$
satisfies all coordinatewise KKT conditions.
\end{corollary}

Corollary $\ref{cor:cd}$ provides a support-dependent threshold above which the OLS fit on a given $m$-variable support is a fixed point of the coordinate-descent updates. This establishes the existence of an $m$-sparse coordinatewise solution, but does not guarantee convergence to that solution or its global optimality. Since the threshold depends on the unknown support $S$, it cannot directly specify an upper endpoint for the regularization search. We instead use the following heuristic value to guide the search: 
we suggest centering and scaling the data, and then computing a crude value $\lambda_{m,\max}$:
$$
\lambda_{m,\max} = \max_{j\notin S} \frac{ |\bx_j|^\top|\by - \bar{\by}| }{ n w_j P_{m-1}(\bg_{-[j]}) }
$$
by plugging in $\wh\bbeta = \epsilon\mathbf{1}_p$ for a small $\epsilon$, such as $0.001$.   Across the synthetic settings, with results presented in Section \ref{sec:sim}, we have verified that this crude value typically results in good estimates.

Each exact coordinate update in \eqref{cdupdate_raw} cannot increase the Fridge objective. Coordinatewise convexity alone, however, does not establish convergence of the iterates or global optimality. For the weighted absolute-value penalty, Lemma \ref{lem:local_convex} gives a sufficient condition under which exact cyclic coordinate descent converges to the unique global minimizer.

\begin{lemma} \label{lem:local_convex}
Fix $1\le m<p$, $\lambda\ge0$, and $g_j(\beta_j)=w_j|\beta_j|$ with $w_j>0$. Let $\bC_n=n^{-1}\bX^\top\bX$ be positive definite, and let $c_*(\bC_n)$ denote its smallest eigenvalue. Write $\hat\bbeta^{\mathrm{OLS}}=(\bX^\top\bX)^{-1}\bX^\top\by$. 
Start exact cyclic coordinate descent \eqref{cdupdate_raw} at $\bbeta^{(0)}$. Choose
$$
r > \left\{\frac{2\bigl(R_{\lambda,m}(\bbeta^{(0)}) - (2n)^{-1}\|\by-\bX\hat\bbeta^{\mathrm{OLS}}\|_2^2 \bigr)}{c_*(\bC_n)}\right\}^{1/2}.
$$
Set $b_j=w_j(|\hat\beta_j^{\mathrm{OLS}}|+r)$ and, with $P_{-1}\equiv1$, define
$
M_r=\max_j\sum_{k\ne j}w_jw_kP_{m-2}(\bb_{-[j,k]}).
$
If $\lambda M_r<c_*(\bC_n)$, the CD iterates converge to the unique global minimizer of $R_{\lambda,m}$.
\end{lemma}
For $m=1$, $M_r = \max_j w_j\sum_{k\ne j}w_k$ is independent of $r$. Hence $\lambda\max_j w_j\sum_{k\ne j}w_k<c_*(\bC_n)$ guarantees convergence of cyclic coordinate descent to the unique global minimizer from any starting point.

\begin{remark}
	We provide an alternative algorithm to find a candidate for a Fridge model in Appendix \ref{sec:irr} that utilizes a local quadratic approximation.
\end{remark}

\subsection{Solution Path Evaluation}  \label{sec:solpathalgo}
Because the Fridge penalty creates a non-linear and potentially multi-modal objective surface, direct optimization at a single, arbitrary $\lambda$ is prone to becoming trapped in poor local minima. 
Instead, by tracing the solution path from a known, well-behaved initial estimate, we can guide the estimator along the solution path.
A straightforward approach to constructing the solution path is to employ the model-fitting algorithms described previously via a {warm-start} strategy. 
Null-shrinkage methods usually start with a large value of $\lambda$ that yields a null model and proceed iteratively with smaller values of $\lambda$ \tred{\citep[Section 2.5]{friedman2010regularization}}.
For Fridge, since the best subset model associated with large $\lambda$ is unknown, we do the opposite.
By starting with a small penalty $\lambda$ (which gives almost OLS estimator) and using the resulting coefficients as initial estimates for the subsequent, slightly larger $\lambda$, one can iteratively trace the path. For regularized generalized linear models, dedicated path algorithms can reduce the number of optimization steps needed to approximate a solution path. This motivates developing a path following approach for Fridge.

\tred{Predictor--corrector methods for regularization paths are developed by \citet{park2007l1}, and differential equation path following by \citet{zhou2014generic}.} An alternative approach treats the coefficient vector $\beta$ as a differentiable function of $\lambda$ for fixed $m$. Write $\mathcal P_m(\bbeta)=P_m(\bg(\bbeta))$ and take its gradient and Hessian with respect to $\bbeta$.
On a smooth stationary branch where the Hessian is nonsingular, the first-order condition is $\bX^\top \bX\bbeta-\bX^\top\by+n\lambda\nabla\mathcal P_m(\bbeta)=0$,
thus implicit differentiation gives
\begin{align}
	\frac{d\bbeta}{d\lambda}
	=-(\bX^\top\bX+n\lambda\nabla^2\mathcal P_m(\bbeta))^{-1}
	 n\nabla\mathcal P_m(\bbeta),
\end{align}
which we leverage the current estimate of $\bbeta_m(\lambda)$ to project the state of the model at the next infinitesimal increment of $\lambda$.

A global minimizer may jump as $\lambda$ varies, so the derivative above describes only a smooth stationary branch. For a numerical predictor direction, replace its penalty Hessian by the nonnegative diagonal matrix $\tilde{\bH}=\operatorname{diag}(|\nabla_j\mathcal P_m(\bbeta)|/|\beta_j|)$ at points with nonzero coordinates. At $\bbeta^{(0)}$, this corresponds to the local quadratic model
\begin{align}
	\mathcal P_m(\bbeta^{(0)})
	+\frac12\sum_{j=1}^p
	\frac{|\nabla_j\mathcal P_m(\bbeta^{(0)})|}{|\beta_j^{(0)}|}
	\{\beta_j^2-(\beta_j^{(0)})^2\},
\end{align}
whose value and gradient match $\mathcal P_m$ at $\bbeta^{(0)}$.
This construction follows the local quadratic approximation principle of \citet{fan2001variable}, whose connection to MM algorithms and Newton-type updates is developed by \citet{hunter2005variable}. Here the diagonal weights depend on the full coefficient vector through the nonseparable Fridge penalty, and we use this approximation within a predictor--corrector scheme to trace the solution path.
Replacing the exact Hessian with $\tilde{\bH}$ gives the predictor direction
\begin{align}
	\frac{d\bbeta}{d\lambda}
	=-(\bX^\top\bX + n\lambda\tilde{\bH})^{-1}n\nabla \mathcal P_m(\bbeta).
\end{align}
By applying a small step $\delta$ in the direction of this approximate derivative, we obtain an initial estimate $\tilde\bbeta_m(\lambda^{\textrm{new}})$ for the subsequent regularization level $\lambda^{\textrm{new}}$.

Damped quasi-Newton corrections then seek to reduce the residual at the stationary condition, again using $\tilde{\bH}$:
\begin{align}
	\bbeta\leftarrow\bbeta-
	\gamma_m(\bX^\top\bX+n\lambda^{\textrm{new}}\tilde{\bH})^{-1}
	 R_m(\bbeta;\lambda^{\textrm{new}}),
\end{align}
where $\gamma_m \in \mathbb{R}$ is the adaptive damping factor, and $R_m(\bbeta;\lambda) = \bX^\top\bX\bbeta - \bX^\top\by + n\lambda\nabla\mathcal P_m(\bbeta)$ is the first-order Karush-Kuhn-Tucker condition.
Starting from $\tilde\bbeta_m(\lambda^{\textrm{new}})$, apply this correction $k$ times to reduce the solution's departure from the stationary condition, recomputing $\tilde{\bH}$ and $R_m$ at the current $\bbeta$ before each step.
We use the heuristic $\gamma_m=\{1\wedge4/(m+1)\}\vee0.1$ to moderate the correction for large $m$, and $k=2$.
Algorithm \ref{algo:solution_path} describes the predictor--corrector process with $g_j(b) = |b|/|\hat\beta_j^\textrm{init}|$.

We note that the diagonal quadratic model is not generally a majorizer, so the method has no MM descent guarantee and need not find a global minimizer. Fits should be checked against the original objective and its stationary conditions. However, this approach performs well empirically, as shown in the subsequent simulation studies and applications.

\begin{algorithm}
\setstretch{1.2}
\caption{Fridge solution path evaluation} \label{algo:solution_path}
\begin{algorithmic}[1]
    \State \textbf{Initialize}
    \State $m$: Integer target model size.
    \State $\delta > 0$: Step size for $\lambda$.
    \State $k\ge1$: Number of correction steps per value of $\lambda$.
    \State $\gamma_m = \{1\wedge4/(m+1)\}\vee 0.1$: Damping factor.
    \State $\hat\bbeta^\text{init} = \hat{\bbeta}^\text{ridge}$: Ridge estimate for adaptive weight.
    \State \textbf{Start}
	\State $\lambda \gets 0$: Initial value of regularization parameter. $0$ corresponds to OLS.
	\State $\bbeta \gets \hat{\bbeta}^{\text{OLS}}$: The Fridge estimate at initial point $\lambda=0$.
	\Repeat 
	  \State $\tilde{\bH} \gets \text{diag}(|\nabla_j\mathcal P_m(\bbeta)|/|\beta_j|)$.  \Comment{Update Step}
	  \State ${d\bbeta}/{d\lambda} \gets - (\bX^\top \bX + n\lambda \tilde{\bH})^{-1}n\nabla\mathcal P_m(\bbeta)$
	  \State $\bbeta \gets \bbeta + \delta \cdot {d\bbeta}/{d\lambda}.$
	  \State $\lambda \gets \lambda + \delta$
	  \For{$1,\ldots,k$} \Comment{Correction steps}
	    \State $\tilde{\bH}\gets\operatorname{diag}(|\nabla_j\mathcal P_m(\bbeta)|/|\beta_j|)$
	    \State $R \gets \bX^\top \bX \bbeta - \bX^\top \by + n\lambda \nabla\mathcal P_m(\bbeta)$
	    \State $\bbeta \gets \bbeta - \gamma_m (\bX^\top \bX+n\lambda \tilde{\bH})^{-1}R$
	  \EndFor
	\Until Target model size is achieved.
    \State \Return $\bbeta$
\end{algorithmic}
\end{algorithm}

\section{Properties}  \label{sec:theory}
We assume the data are generated by $\by_n = \bX_n\bbeta_0 + \boldsymbol{\varepsilon}$ where ${\varepsilon}_{i}$'s are independent normal random variables with mean $0$ and variance $\sigma^2$, unless otherwise specified. Throughout this section, we assume $\bX_n$ is deterministic.
Suppose, without loss of generality, the true coefficient be $\bbeta_0 = (\bbeta_{10}^T, \bbeta_{20}^T)^T$, where $\bbeta_{10}$ is the active set coefficients (nonzero elements; $\mathcal{A}$) of size $q$, and $\bbeta_{20} = \bzero$ is the inactive set ($\mathcal{A}^c$) coefficients with size $p-q$. 
Throughout this section, we assume that $n > p$ and $n^{-1}\bX_n^\top\bX_n$ is invertible and converges to a positive definite matrix $\bC$ as $n\to\infty$. Specifically, we denote 
$$
\setstretch{0.9}
\bC = \begin{bmatrix} 
	\bC_{11} & \bC_{12} \\
	\bC_{21} & \bC_{22}
\end{bmatrix}
$$
so that $\bC_{11}$ corresponds to $\bbeta_{10}$.

Throughout this section, we assume that $g_j(b) = |b|/|\hat\beta_j^\text{init}|$, where $\hat\bbeta^\text{init}$ is a $\sqrt{n}$-consistent initial estimator, i.e., $\hat{\bbeta}^\text{init} - \bbeta_0 = O_p(n^{-1/2})$, and that $0<|\hat\beta_j^{\mathrm{init}}|<\infty$ for all $j$. We focus our analysis on this scaled formulation, because in practice, adaptive weighting stabilizes the optimization landscape by scaling intermediate solutions close to one. Appendix \ref{A:stdfridge} presents results for Fridge with $g_j(b_j) = |b_j|$, i.e., a version without adaptive weighting.

\subsection{Limiting Distribution}
Following the theory of \citet{knight2000asymptotics}, with a standard rate of $\lambda_n$, we have asymptotic normality of the Fridge estimator.

{\small
\begin{theorem}[Limiting distribution] \label{thm:limiting}
Let $R_n(\bbeta) = (2n)^{-1}\|\by_n-\bX_n\bbeta\|_2^2 + \lambda_nP_m(\bbeta)$. Assume that
the adaptive weights are finite with probability tending to one. Suppose
$
\sqrt n\,\lambda_n\to\lambda_0\ge0.
$
If $m\le q$, assume additionally that
$
n\lambda_n\to\infty.
$
If $m>q$, assume that $\bC_{12}=\bzero$.
Let
$
\hat\bbeta
=
(\hat\bbeta_1^\top,\hat\bbeta_2^\top)^\top
$
be a global minimizer of $R_n(\bbeta)$. Then
$$
\sqrt n(\hat\bbeta_1-\bbeta_{10})
\xrightarrow{d}
\mathcal N_q
\left(
  -\lambda_0\bC_{11}^{-1}\bd_{10}, \;
  \sigma^2\bC_{11}^{-1}
\right),
$$
where
$$
(\bd_{10})_j
=
\frac{1}{\beta_{0j}}
\binom{q-1}{m},
\qquad j\in\mathcal A,
$$
with the convention $\binom{q-1}{m}=0$ for $m>q-1$.
\end{theorem}
}

\begin{corollary}  \label{cor:oracle}
	Under the assumptions of Theorem \ref{thm:limiting},
	\begin{enumerate}
		\item[(i)] If $m \ge q$, then $\bd_{10} = \bzero$, and the active set coefficient estimators are asymptotically unbiased.
		\item[(ii)] If $m \le q$ and $n\lambda_n \to \infty$, then $\pr(\hat{\bbeta}_2 = \bzero) \to 1$, thus the Fridge estimator screens out all inactive variables asymptotically.
	\end{enumerate}
\end{corollary}

Corollary implies that whenever $m \ge q$, all true active variables are retained with probability tending to one, and their coefficient estimates have $O_p(n^{-1/2})$ estimation error. 
When $m \le q$, Fridge filters out all irrelevant variables from the model with probability tending to one. These results imply that at $m=q$, the Fridge estimator consistently recovers the true support and has the same centered asymptotic normal distribution for the active coefficients as the oracle least-squares estimator, thereby attaining the oracle property.

\subsection{Best Subset Selection Property}

In Section \ref{sec:def}, we introduced Fridge as a model that bridges regularized regression and best subset selection. The following result shows that,
for a fixed sample, increasing the regularization parameter indeed drives global Fridge minimizers toward the set of best subset least-squares solutions with at most $m$ nonzero coefficients. If the best subset solution is unique, the Fridge estimator converges to it as $\lambda\to\infty$.

\begin{theorem}  \label{thm:bss}
Fix $n$ and $m<p$. Let $R_{\lambda,m}(\bbeta)=\frac{1}{2n}\|\by_n-\bX_n\bbeta\|_2^2+\lambda P_m(\bg(\bbeta))$ and $\hat\bbeta_m(\lambda)$ be a global minimizer. Let $\hat\bbeta_m^{\mathrm{BSS}}\in\arg\min_{\|\bbeta\|_0\le m}\frac{1}{2n}\|\by_n-\bX_n\bbeta\|_2^2.$
Then every accumulation point of $\hat\bbeta_m(\lambda)$ as $\lambda\to\infty$ is a best subset minimizer. If the best subset minimizer is unique, then $\hat\bbeta_m(\lambda)\to\hat\bbeta_m^{\mathrm{BSS}}.$
\end{theorem}  

While technically it is possible to fit a Fridge model with true model size for best subset selection purpose, our primary interest is in identifying a model that balances prediction and generalizability, and not in the recovery of the best subset selection. We delegate further study of the optimal best subset selection via Fridge to Appendix \ref{sec:bss_via_fridge}.

\section{Numerical Study}   \label{sec:sim}

We conduct a series of simulation studies to evaluate Fridge's behavior and performance compared to existing null-shrinkage and best subset selection methods. Our setting follows the settings proposed in \cite{hastie2020best}, which rigorously compares methods in terms of F-score, proportion of variance explained (PVE), relative test error (RTE), and sparsity of the solution, under varying signal-to-noise ratios (SNR) and predictor correlation.

We tested the following combinations of  sample size ($n$), number of predictors ($p$), and true model size ($q$): $(n,p,q) = (100, 10, 5), (500, 100, 5), (100, 1000, 10), (50, 1000, 5)$.
%
We considered ten values for the SNR ranging from 0.05 to 6 on a log scale, and three values ($\rho = 0, 0.35, 0.7$) for the correlation between the adjacent predictors. Magnitudes of the true regression coefficients were chosen to be $1$ with randomly chosen signs. We chose the step, i.e., the index difference between the nearest located true non-zero predictors, so that true non-zero coefficients are either adjacent to each other (Step $=1$) or occur at equally spaced indices with a single true zero coefficient between them (Step $=2$). {Details about the data generating process} and evaluation metrics are provided in Appendix \ref{A:sim_detail}.

We tested the proposed Fridge regression aslong with the Lasso \citep{tibshirani1996regression}, adaptive Lasso \citep{zou2006adaptive}, non-convex penalties \citep{fan2001variable, zhang2010nearly}, and best subset selection \citep{zhu2020polynomial}. For the adaptive Lasso, initial weights are computed via ridge regression with penalty parameter chosen from a 10-fold cross-validation. For the initial weights of Fridge, we use ridge coefficients with penalty parameter $10^{-6}/m$, where the target model size $m$ is selected from the set $\{1, 2, 3, 4, 5, 8, 10, 15\} \cap [1\dots p]$ along with $\lambda$ using the 10-fold cross-validation. We deliberately exclude $0$ from the candidate values to prevent Fridge from reducing to the adaptive Lasso.

\begin{figure}
\centering
  \includegraphics[width=0.8\textwidth]{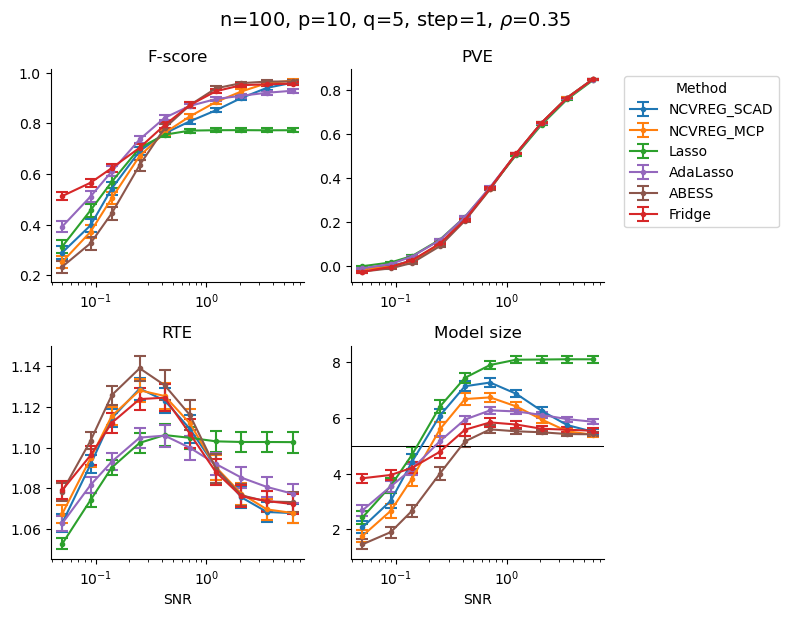}
  \caption[F-score, PVE, RTE, and model size in the setting with $n=100$, $p=10$, step $=1$, and $\rho=0.35$.]{
    F-score, PVE, RTE, and model size for $n=100$, $p=10$, Step $=1$, and $\rho=0.35$. The horizontal axis shows the SNR values ranging from $0.05$ to $6$, equally spaced on a $\log_{10}$ scale. Points show means over 500 independent Monte Carlo replications; error bars indicate $\pm 2$ standard errors. The solid black line in the model size plot indicates the true model size $q=5$.
  }
  \label{fig:step1_rho0.35_n100_p10}
\end{figure}

Figure \ref{fig:step1_rho0.35_n100_p10} displays the results of 500 independent Monte Carlo replications for the setting $(n,p,q)=(100,10,5)$, $\text{Step}=1$, and $\rho=0.35$. Full results for all configurations, each averaged over 500 independent trials, are provided in Appendix \ref{A:sim_detail}, Figures \ref{fig:step1_fscore}--\ref{fig:step2_modelsize}. Across all plots, the horizontal axis (SNR) is logarithmically scaled with base $10$, while the vertical axes (F-score, PVE, RTE, or model size) are linear.

Across the simulation configurations, the Fridge estimator delivers competitive or superior metrics throughout the range of SNRs. Specifically, Fridge consistently achieves high F-scores across all evaluated conditions, whereas competing benchmark methods excel only within specific settings or specific ranges of SNRs.
With the exception of the adaptive Lasso, all models achieve similar PVE across configurations. Specifically, when $n < p$, the adaptive Lasso explains significantly less variance than the remaining methods.
Under low-SNR conditions, the Lasso and the adaptive Lasso outperformed other methods in terms of RTE. This pattern is reversed under high-SNR regime, where the Lasso and the adaptive Lasso fail to reduce the testing error.

When $n > p$, Fridge consistently selects a model size closer to the true $q$ across varying signal-to-noise ratios. When $n < p$ and noise dominates the signal, e.g., SNR $< 0.5$, Fridge tends to retain a large number of variables (Figures \ref{fig:step1_modelsize}, \ref{fig:step2_modelsize}). While the Fridge models contain extra non-zero entries, an analysis of the coefficient magnitudes demonstrates a soft-thresholding effect that suppresses noise variable variance.
In the tested low-signal, low-evidence environments, forcing strict sparsity may force the model to select spurious noise variables, which can lead to poor generalization. 
Fridge instead assigns many small, non-zero coefficients rather than forcing aggressive truncation. For instance, in a setting with $(n,p,q) = (50,1000,5)$, only 17 coefficients exceeded $0.05$, and only 7 were larger than $0.1$ in absolute value. This demonstrates that Fridge transitions to a ridge-like model under low-SNR conditions, spreading weights across many variables to minimize the damage caused by selecting any spurious covariates. The corresponding F-score, PVE, and RTE show that Fridge remains one of the best-performing models, supporting its ridge-like transitioning behavior.

As the true signal becomes more pronounced, Fridge adapts, filters out irrelevant variables, and restores sparsity without requiring manual algorithmic switches.
This ability to transition between sparse and dense models acts as a safety net, prioritizing prediction stability over arbitrary sparsity when the true variables cannot be reliably identified. 

\section{Riboflavin Production Data} \label{sec:ribo}
We examined the riboflavin production data of \citet{buhlmann2014biology}, which contain $n=71$ observations, $p=4088$ gene-expression measurements from \textit{Bacillus subtilis}, and a log-transformed production response. We fitted Fridge using the solution path implementation (Section \ref{sec:solpathalgo}) and selected its target model size and regularization parameter by 10-fold cross-validation with the one-standard-error rule \citep[Section 7.10]{hastie2009elements}. This gave $m=2$ and $\lambda\approx1.72\times10^{-4}$. For comparison, we fitted 10-fold cross-validated SCAD and a cross-validated adaptive Lasso initialized by cross-validated ridge regression. We also fitted SuSiE with a maximum of 34 effects \citep{wang2020susie}; its posterior inclusion probabilities (PIPs) and credible set serve as meaningful descriptive references.

\begin{figure}[!htp]
\centering
\includegraphics[width=1.0\textwidth]{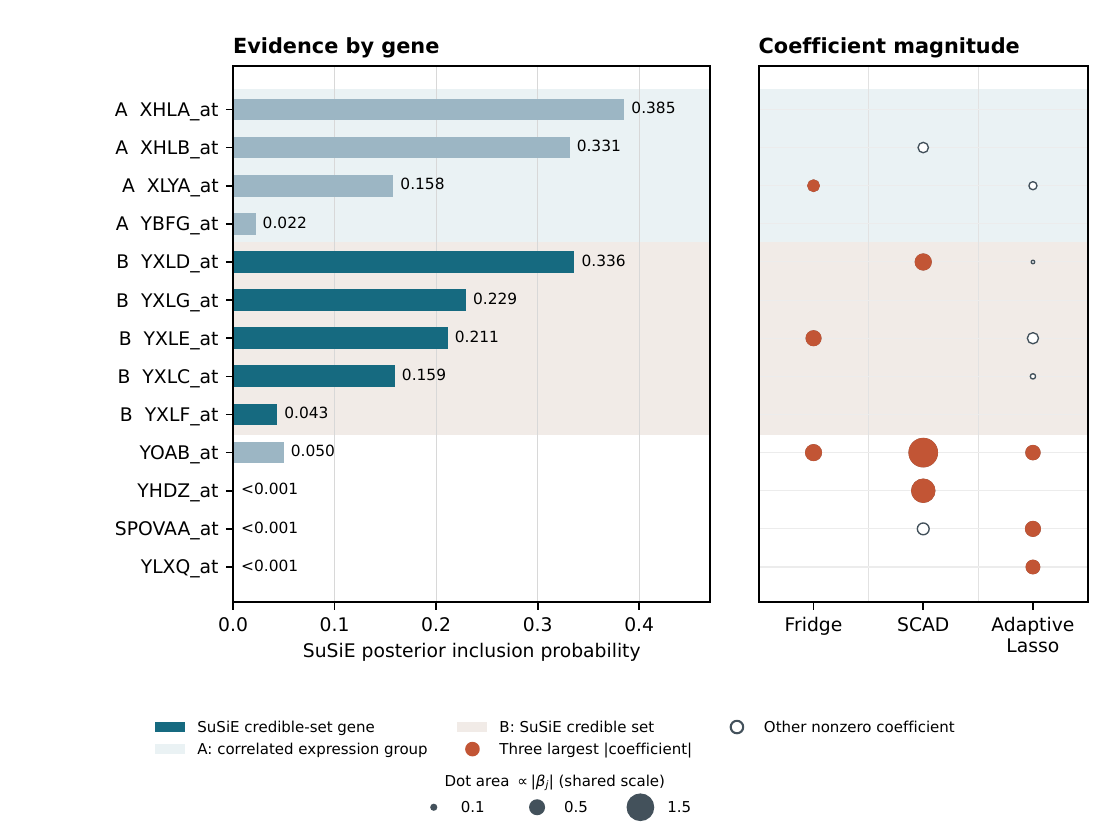}
\caption{Gene-level comparison for the ten highest SuSiE PIPs and any additional genes among a method's three largest absolute coefficients. Shaded rows and letter prefixes identify two correlated expression groups: A is defined from sample correlations and B is SuSiE's sole reported credible set. Within each group, genes are ordered by PIP. Dark bars indicate credible-set membership. Filled circles mark each method's three largest absolute coefficients; open circles indicate other nonzero coefficients. Circle area is proportional to $|\beta_j|$, using a common scale across methods.}
\label{fig:ribo_pip}
\end{figure}

The fitted Fridge model had 8 nonzero coefficients, compared with 16 for SCAD and 53 for adaptive Lasso. SuSiE returned one five-gene credible set (group B in Figure \ref{fig:ribo_pip}), comprising \texttt{YXLC\_at}, \texttt{YXLD\_at}, \texttt{YXLE\_at}, \texttt{YXLF\_at}, and \texttt{YXLG\_at}. Its members have pairwise sample correlations between $0.969$ and $0.986$. A separate empirical correlation group (A), which is not a SuSiE credible set, contains \texttt{XHLA\_at}, \texttt{XHLB\_at}, \texttt{XLYA\_at}, and \texttt{YBFG\_at}, with pairwise correlations between $0.914$ and $0.985$. Fridge selects one representative from each group, \texttt{XLYA\_at} and \texttt{YXLE\_at}; SCAD likewise selects one from each, whereas adaptive Lasso selects one from A and three from B. Among all selected genes, Fridge and SCAD contain no pairs with absolute sample correlation above $0.9$; adaptive Lasso contains four such pairs. Among 500 bootstrap Fridge refits at fixed $m$ and $\lambda$, at least one member of B was selected in 97.6\% of samples, although \texttt{YXLE\_at} itself was selected in 60.6\%.

To compare equally-sized sets, Figure \ref{fig:ribo_pip} marks the three largest absolute coefficients from each method next to SuSiE's PIPs. All three leading Fridge genes (\texttt{YOAB\_at}, \texttt{YXLE\_at}, and \texttt{XLYA\_at}) are among the ten highest-PIP genes, versus two of SCAD's and one of adaptive Lasso's. Across the full selected sets (Table \ref{tab:ribo_concordance}), Fridge selects 3 of these high-PIP genes among 8 nonzero coefficients, leaving 5 outside the list. SCAD also selects 3, but among 16 coefficients, leaving 13 outside; adaptive Lasso selects 5 among 53, leaving 48 outside. This behavior coincides with Fridge's high F-score in Section \ref{sec:sim}.
Fridge has fewer selections discordant with SuSiE's high-PIP ranking.

\begin{table}[!htp]
\centering
\caption{Concordance with SuSiE's ten highest-PIP genes. ``In top 10'' compares each method's full set of nonzero coefficients.  ``Leading 3 in top 10'' compares equal-sized sets. The numbers in parentheses indicates the percentage of all or leading 3 nonzero variables that coincide with the top-10 PIP variables. The final column counts all nonzero coefficients in SuSiE's credible set. The numbers in parentheses indicates the percentage of all or nonzero variables that coincide with the credible set.}
\label{tab:ribo_concordance}
\small
\begin{tabular}{lrrrrr}
\hline
Method & Non-zero & In top 10 & Leading 3 in top 10 & In credible set \\
\hline
Fridge & 8 & 3 (37.5\%) & 3 (100\%) & 1 (12.5\%) \\
SCAD & 16 & 3 (18.8\%) & 2 (66.7\%) & 1 (6.3\%) \\
Adaptive Lasso & 53 & 5 (9.4\%) & 1 (33.3\%) & 3 (5.7\%) \\
\hline
\end{tabular}
\end{table}

%
%
\section{Discussion}  \label{sec:discussion}

Fridge provides a way to specify the model size toward which regularization is directed. The target model size $m$ determines where the penalty vanishes, while $\lambda$ controls the strength of shrinkage toward that set of models. This distinction allows selective shrinkage at finite regularization levels and gives a connection to best subset selection under extreme regularization. The target model size $m$ therefore has a direct interpretation, although the number of predictors retained in a fitted model can exceed $m$. In applications focused on prediction, both parameters can be selected by cross-validation, as in the riboflavin analysis.

The simulations suggest that this flexibility is useful when the evidence for individual predictors varies with signal strength. Fridge produced competitive prediction and selection results across the settings studied. Under weak signals in high dimensions, it often retained many small coefficients; with stronger signals, its fitted models became sparse. In alignment with \citet{hastie2020best}, these results suggest that allowing small coefficients to persist can help preserve predictive information when exact variable selection is difficult. 

Computation and tuning remain central limitations. The Fridge objective is generally nonconvex, and the solution path algorithm follows a stationary branch that may differ from the path of global minimizers. The sufficient condition in Lemma \ref{lem:local_convex} gives a setting in which coordinate descent converges to the global solution, but the general algorithms do not carry that guarantee. Objective comparisons across starting values and checks of the stationary conditions are therefore useful in practice. Likewise, the oracle property concerns global minimizers under specified assumptions and tuning rates; it does not establish the oracle property for a cross-validated numerical fit. However, the oracle properties of the competing methods likewise require specific assumptions and tuning rates, and therefore do not automatically apply to their cross-validated numerical fits.

The riboflavin application illustrates the potential for parsimonious selection in biological studies. Fridge selected fewer genes outside SuSiE's ten highest-PIP genes than SCAD or the adaptive Lasso, and all three of its leading coefficients belonged to that list. It also selected one representative from each of the two reported correlation groups. Agreement with SuSiE provides a useful comparison, but we note that this does not establish biological relevance or identify false discoveries. The bootstrap results further distinguish stability of a group from stability of an individual gene: selection of at least one credible set member was substantially more stable than selection of the particular gene chosen in the original fit. Independent data or experimental validation would be needed to assess the biological interpretation.

Further work on Fridge could develop more reliable optimization methods, characterize sensitivity to the initial weights, and establish prediction and selection guarantees for finite regularization when the number of predictors grows with the sample size. Extensions to generalized linear models and structured groups of predictors would broaden the scope of the method. In applications with strongly correlated features, reporting stability at the group level alongside coefficient estimates may also provide a more informative account of what Fridge has identified.

  \bibliography{bibliography.bib}

\newpage

\appendix
\section{Proof of Lemma \ref{lem:local_convex}}
Write $R=R_{\lambda,m}$, $c=c_*(\bC_n)$ for simplicity. The least-squares decomposition gives
\begin{align*}
R(\bbeta) 
&\ge \tfrac1{2n}\|\by-\bX\hat\bbeta^{\mathrm{OLS}}\|_2^2 + \tfrac12(\bbeta - \hat\bbeta^\text{OLS})^\top \bC_n (\bbeta - \hat\bbeta^\text{OLS}) \\
&\ge \tfrac1{2n}\|\by-\bX\hat\bbeta^{\mathrm{OLS}}\|_2^2 +
\tfrac c2\|\bbeta-\hat\bbeta^{\mathrm{OLS}}\|_2^2.    
\end{align*}
Thus, for the chosen $r$, $K :=\{\bbeta:R(\bbeta)\le R(\bbeta^{(0)})\}$ is a subset of $B_r := \{\bbeta:\|\bbeta-\hat\bbeta^{\mathrm{OLS}}\|_2<r\}$. It follows that $R$ is coercive, and the compact sublevel set $K$ contains every iterate and every global minimizer.

On $B_r$, suppose $\bbeta$ includes no zero coordinates. We have $w_j |\beta_j| \le w_j(|\hat\beta_j^\text{OLS}| + r) = b_j$. The penalty Hessian has zero diagonal and off-diagonal absolute entries $w_jw_kP_{m-2}(\bg_{-[j,k]})$, and thus its operator norm is at most $M_r$. Hence, $\nabla^2R(\bbeta) \succeq (c-\lambda M_r) I$.

Now, to include zero coordinates, take any $\bbeta(t) = \mathbf u+t\mathbf v$ in $B_r$ and set $f(t) = R(\bbeta(t))-\frac{c-\lambda M_r}2\|\bbeta(t)\|_2^2$. Coordinates that vanish identically along this line segment can be omitted, since every penalty term containing them is zero. Between the finitely many remaining zero crossings, the same Hessian bound gives $f''(t)\ge0$. At a crossing $t_0$, let $J=\{j:\beta_j(t_0)=0,\ v_j\ne0\}$, where $v_j$ is the $j$th component of $\bv$, and $\bg_0=\bg(\bbeta(t_0))$. The one-sided derivatives satisfy
$$
f'_+(t_0)-f'_-(t_0)
=2\lambda\sum_{j\in J}w_j|v_j|P_{m-1}((\bg_0)_{-[j]})\ge0,
$$
since $w_j|\beta_j(t)|$ has slope jump $2w_j|v_j|$. In addition, products with two or more vanishing factors have zero derivative at $t_0$. Thus, $f$ has nondecreasing slope along every line segment, proving that $R$ is $(c-\lambda M_r)$-strongly convex on $B_r$.

Indexing individual coordinate updates by $t$, strong convexity of each coordinate subproblem gives
$
R(\bbeta^{(t)})-R(\bbeta^{(t+1)})
\ge\tfrac c2\|\bbeta^{(t+1)}-\bbeta^{(t)}\|_2^2.
$
Summing these inequalities, the total objective decrease is finite, thus $\sum_t \|\bbeta^{(t+1)} - \bbeta^{(t)}\|_2^2 < \infty$. Thus, successive changes vanish. 

All CD iterates lie in  \(K\), which is compact, thus accumulation points exist. Because updates are cyclic, successive changes vanish, and since the soft-thresholding update is continuous, every accumulation point minimizes \(R\) with respect to each coordinate separately.
Since the first directional derivative of $R$ is the sum of its coordinatewise directional derivatives, it is nonnegative in every direction at such a point by coordinate-wise optimality. By strong convexity, this point is the unique minimizer on $B_r$. Since every global minimizer lies in $K\subset B_r$, the whole sequence converges to this unique global minimizer. \qed

\section{Iteratively Reweighted Regression}  \label{sec:irr}
As discussed in Section \ref{sec:weighted}
the Fridge penalty can be seen as an adaptively weighted penalty. 
Here, we assume $g_j(\beta_j)=|\beta_j|$ and write $v_j(\bg)=P_{m-1}(\bg_{-[j]})=P_{m-1}(\bg)-\bV_{m-1}(\bg)[j]$. Counting each product once for each of its $m+1$ factors gives the identity
\begin{align}
\label{eq:dd2}
\lambda P_m(\bg)
&=\frac{\lambda}{m+1}\sum_{j=1}^p v_j(\bg)|\beta_j|.
\end{align}
This identity also holds when some coefficients are zero.

To construct an iteratively reweighted Lasso (IRL) step, use the first-order penalty approximation at the current iterate $\wh\bbeta^{(t)}$, with $v_j^{(t)}=v_j(|\wh\bbeta^{(t)}|)$:
\begin{align}
\label{eq:dd3}
\wh\bbeta^{(t+1)}\in\arg\min_{\bbeta}
\left\{\frac{Q(\bbeta)}2+\lambda\sum_{j=1}^p v_j^{(t)}|\beta_j|\right\}.
\end{align}
An additive constant, $\lambda\{P_m(|\wh\bbeta^{(t)}|)-\sum_jv_j^{(t)}|\wh\beta_j^{(t)}|\}$, makes the approximation agree with the original objective at the current iterate but does not affect the minimizer. The factor $1/(m+1)$ from \eqref{eq:dd2} is absent from \eqref{eq:dd3}: at a fixed point, the weighted-Lasso KKT equations then match the stationary conditions of $Q/2+\lambda P_m(|\bbeta|)$. This first-order approximation is not generally a majorizer, so the iteration has no global optimality guarantee.

The weight $v_j$ can be exactly zero, thus ensuring that a portion of coefficients are not constrained, which happens when $P_{m-1}(\bg) = \bV_{m-1}(\bg)[j]$ and $g(x)=|x|$. For example, when the TMS $=1$, $v_j$ is zero if $\sum_{l\ne j}|\beta_l|$ is zero, and when TMS $=2$, if $\sum_{\substack{l<m, \\ l,m \ne j}}|\beta_l\beta_m| = 0$. These scenarios imply $\beta_j$ should be among the allowable non-zero coefficients as the other $(p-m)$ coefficient estimates are already set to zero. Thus the Fridge will not actively penalize the remaining $\beta_j$ in the subsequent process.

A local quadratic alternative is the weighted-ridge step
$$
\wh\bbeta^{(t+1)}\in\arg\min_{\bbeta}
\left\{\frac{Q(\bbeta)}2+
\frac\lambda2\sum_{j=1}^p\frac{v_j^{(t)}}{|\wh\beta_j^{(t)}|}\beta_j^2\right\}.
$$
Its gradient matches the Fridge penalty gradient at the current iterate. This iteratively reweighted ridge (IRR) step requires a specified treatment of zero coefficients and likewise has no general majorization guarantee.

IRL uses a local linear penalty approximation\tred{, analogous to the approach of \citet{zou2008one}}; IRR uses a local quadratic one as in \citet{fan2001variable}.
The inner weighted Lasso and weighted ridge problems can use existing solvers, including suitable screening methods for the Lasso step \citep{el2010safe, tibshirani2012strong}. Similar inner problems can be formed for generalized linear models. Their solutions remain numerical candidates for the Fridge objective.

\section{Proof of Theorem \ref{thm:limiting}}  \label{proof:thm2}

Following \citet{knight2000asymptotics}, we minimize the shifted local objective function. Let $\hat \bw = (\hat w_j) = (1/|\hat\beta_j^\textrm{init}|)$ be the adaptive weight for simplicity. Write
$
\bW_n=\frac{1}{\sqrt n}\bX_n^\top\boldsymbol\varepsilon_n.
$
By assumption,
$
\bW_n\xrightarrow{d}\mathcal N_p(\bzero,\sigma^2\bC).
$
Also note that
$
n^{-1}\bX_n^\top\boldsymbol\varepsilon_n=O_p(n^{-1/2}).
$
We consider separately the cases $m\le q$ and $m>q$.

\medskip
\noindent
\textbf{Case 1: $m\le q$.}

We first establish consistency of the global minimizer. Let
$
\bdelta_n := \hat\bbeta-\bbeta_0.
$
Since $R_n(\hat\bbeta)\le R_n(\bbeta_0)$ and $P_m(\hat\bbeta)\ge0$, we have
$$
\frac12\bdelta_n^\top\bC_n\bdelta_n
-
\frac{1}{n}
\boldsymbol\varepsilon_n^\top\bX_n\bdelta_n
\le
\lambda_nP_m(\bbeta_0).
$$

For active coordinates,
$
\hat w_j \xrightarrow{p} |\beta_{0j}|^{-1},
$
and hence $P_m(\bbeta_0)=O_p(1)$. Since
$\lambda_n=O(n^{-1/2})$, positive definiteness of $\bC_n$ gives
$
\|\bdelta_n\|_2=O_p(n^{-1/4}) = o_p(1).
$
Thus every global minimizer lies, with probability tending to one, in a fixed neighborhood of $\bbeta_0$ in which all active coefficients remain bounded away from zero.

Now restrict the criterion to $\bbeta_2=\bzero$, and let
$
\tilde\bbeta_1
\in
\argmin_{\bbeta_1}
R_n(\bbeta_1,\bzero).
$
Similar to the previous argument, we can show that $\|\tilde\bbeta_1 - \bbeta_{10}\|_2 = O_p(n^{-1/4}) = O_p(1)$.
On a neighborhood of $\bbeta_{10}$, the signs of the active coefficients are fixed, so the restricted penalty is smooth. Its Hessian is bounded in probability, whereas $\lambda_n\to0$. Therefore the restricted criterion is strictly convex in that neighborhood with probability tending to one.

The first-order condition for $\tilde\bbeta_1$ is given as
$$
\bC_{11}^n
(\tilde\bbeta_1-\bbeta_{10})
-
\frac{1}{n}\bX_{n,1}^\top\boldsymbol\varepsilon_n
+
\lambda_n\bd_n(\tilde\bbeta_1)
=
\bzero,
$$
where
$
\bd_n(\bbeta_1)
=
\nabla_{\bbeta_1}P_m(\bbeta_1,\bzero).
$
Thus, we have
$$
\tilde\bbeta_1-\bbeta_{10}
=
O_p(n^{-1/2}+\lambda_n)
=
O_p(n^{-1/2}).
$$

For $j\in\mathcal A$, i.e., if the coordinate $j$ is non-zero,
$$
\{\bd_n(\tilde\bbeta_1)\}_j
=
\hat w_j
\operatorname{sign}(\tilde\beta_j)
P_{m-1}(\tilde\bbeta_{\mathcal A\setminus\{j\}}) \xrightarrow{p} (\bd_{10})_j,
$$
since
$
\hat w_j \xrightarrow{p} |\beta_{0j}|^{-1}$ and $g_k(\tilde\beta_k) \xrightarrow{p} 1$ for $k\in\mathcal A$.
For $m=q$, the binomial coefficient is zero and the restricted penalty is identically zero.

We next show that the inactive coordinates are exactly zero with probability tending to one. For $j\in\mathcal A^c$, define the score evaluated at the restricted estimator by
$$
a_{nj} =\frac{1}{n} \bx_j^\top (\by_n-\bX_{n,1}\tilde\bbeta_1).
$$
Since
$\tilde\bbeta_1-\bbeta_{10}=O_p(n^{-1/2})$, we have $\max_{j\in\mathcal A^c}|a_{nj}| = O_p(n^{-1/2})$, where $p$ is fixed.

At $(\tilde\bbeta_1,\bzero)$, the zero-coordinate KKT threshold for an inactive coordinate $j$ equals
$$
t_{nj} := \lambda_n \hat w_j P_{m-1}(\tilde\bbeta_1,\bzero).
$$
Because $m\le q$,
$$
P_{m-1}(\tilde\bbeta_1,\bzero) \xrightarrow{p} \binom qm > 0.
$$

Moreover, note that we can rewrite $\hat w_j = {\sqrt n} / {|\sqrt n\,\hat\beta_j^{\mathrm{init}}|}.$ 
For $j\in\mathcal A^c$,
since
$\sqrt n\,\hat\beta_j^{\mathrm{init}}=O_p(1)$
and $p$ is fixed, for every $\epsilon>0$ there is a finite $M$ such that, with probability at least $1-\epsilon$,
$
\max_{j\in\mathcal A^c}
|\sqrt n\,\hat\beta_j^{\mathrm{init}}|
\le M.
$
On this event, we have
$
\min_{j\in\mathcal A^c}t_{nj}
\ge
c\,\lambda_n\sqrt n
$
for some constant $c>0$. Hence,
$$
\frac{
\min_{j\in\mathcal A^c}t_{nj}
}{
n^{-1/2}
}
\ge
c\,n\lambda_n
\xrightarrow{} \infty,
$$
and it follows that
$$
\pr\left(
|a_{nj}|<t_{nj}
\text{ for every }j\in\mathcal A^c
\right)
\to1.
$$

It remains to verify that these strict inactive KKT inequalities identify the global minimizer. In a fixed neighborhood $N$ of $\bbeta_0$, write
$
F_n(\bbeta_1) := R_n(\bbeta_1,\bzero).
$
Write $N_1=\{b:(b,\bzero)\in N\}$. All derivatives with respect to the active coordinates below are derivatives of the restricted map $b\mapsto P_m(b,\bzero)$.
The restricted Hessian of $F_n$ is then
$
\bC_{11}^n+\lambda_n\bH_n(\bbeta_1),
$
where $\bH_n(\bbeta_1) = \nabla^2_{\bbeta_1}P_m(\bbeta_1,\bzero) = O_p(1)$. Since
$\bC_{11}^n$ is uniformly positive definite and
$\lambda_n\to0$, $F_n$ is uniformly strongly convex in this neighborhood.

Furthermore, for any $(\bbeta_1,\bbeta_2)$ in the same neighborhood,
nonnegativity of every product term gives
$$
P_m(\bbeta_1,\bbeta_2)
-
P_m(\bbeta_1,\bzero)
\ge
P_{m-1}(\bbeta_1)
\sum_{j\in\mathcal A^c}\hat w_j|\beta_j|.
$$

Since $P_{m-1}(\bbeta_1)$ is bounded away from zero here, a Taylor expansion about
$(\tilde\bbeta_1,\bzero)$, together with positive definiteness of
$\bC_n$, yields, for some $c_0,c_1>0$,
\begin{align*}
&R_n(\bbeta_1,\bbeta_2)-R_n(\tilde\bbeta_1,0) \\
&= \tfrac12(\bbeta_1-\tilde\bbeta_1,\bbeta_2)^\top C_n(\bbeta_1-\tilde\bbeta_1,\bbeta_2)-a_n^\top \bbeta_2\\
&\quad+ \lambda_n\{P_m(\tilde\bbeta_1+\bbeta_1-\tilde\bbeta_1,0) - P_m(\tilde\bbeta_1,0)
          -\nabla_{\bbeta_1}P_m(\tilde\bbeta_1,\bzero)^\top (\bbeta_1-\tilde\bbeta_1)\}\\
&\quad+ \lambda_n\{P_m(\bbeta_1,\bbeta_2)-P_m(\bbeta_1, 0)\} \\
&\ge \tfrac12(\bbeta_1-\tilde\bbeta_1,\bbeta_2)^\top C_n(\bbeta_1-\tilde\bbeta_1,\bbeta_2)-a_n^\top \bbeta_2 \\
&\quad- \tfrac{\lambda_n}{2} \sup_{\bb\in N_1} \|\nabla^2_{\bb} P_m(\bb, \bzero)\|_2 \, \|\bbeta_1 - \tilde\bbeta_1\|_2^2
 + \lambda_n c_0 \sum_{j \in \mathcal{A}^c} \hat w_j |\beta_{j}| \\
&\ge c_1
\left\|( \bbeta_1-\tilde\bbeta_1, \, \bbeta_2 )^\top\right\|_2^2
+ \sum_{j\in\mathcal A^c} |\beta_j| \{c_0\lambda_n\hat w_j-|a_{nj}|\}
\end{align*}
with probability tending to one.
The quantities in braces are strictly positive with probability tending to one by the preceding KKT bound. Hence
$
(\tilde\bbeta_1,\bzero)
$
is the unique minimizer in that neighborhood with probability tending to one. Since every global minimizer is already consistent, it follows that $\pr\{ \hat\bbeta=(\tilde\bbeta_1^\top,\bzero^\top)^\top \} \to 1$. 
In particular, $\pr(\hat\bbeta_2=\bzero) \to 1.$

Multiplying the active first-order condition by $\sqrt n$ gives
$$
\sqrt n(\tilde\bbeta_1-\bbeta_{10})
= (\bC_{11}^n)^{-1} \left\{ 
  \bW_{n,1} - \sqrt n\,\lambda_n \bd_n(\tilde\bbeta_1)
\right\} 
\xrightarrow{d} \mathcal N_q \left(
  -\lambda_0\bC_{11}^{-1}\bd_{10}, \;
  \sigma^2\bC_{11}^{-1}
\right)
$$
by Slutsky's theorem.
Notice that no assumption on $\bC_{12}$ is required in this case because
$\hat\bbeta_2=\bzero$ with probability tending to one.

\medskip
\noindent
\textbf{Case 2: $m>q$.}

In this case
$
P_m(\bbeta_0)=0.
$
Let
$
\bdelta_n=\hat\bbeta-\bbeta_0.
$
Global optimality gives
$$
\frac12\bdelta_n^\top\bC_n\bdelta_n
-
\frac1n
\boldsymbol\varepsilon_n^\top\bX_n\bdelta_n
+
\lambda_nP_m(\hat\bbeta)
\le0.
$$

Since $\bC_n$ is uniformly positive definite and
$n^{-1}\bX_n^\top\boldsymbol\varepsilon_n=O_p(n^{-1/2})$,
$
\|\bdelta_n\|_2=O_p(n^{-1/2}).
$
The same inequality also implies
$
\lambda_nP_m(\hat\bbeta)
=
O_p(n^{-1}).
$

Because the active coefficients are bounded away from zero with probability tending to one, $P_m$ is differentiable with respect to every active coordinate. For $j\in\mathcal A$,
$$
\left|
\frac{\partial P_m(\hat\bbeta)}
{\partial\beta_j}
\right|
\le
\frac{P_m(\hat\bbeta)}
{|\hat\beta_j|}.
$$
Consequently,
$
\sqrt n \lambda_n \| \nabla_{\bbeta_1}P_m(\hat\bbeta) \| = o_p(1).
$
The active first-order condition is
$$
\bC_{11}^n(\hat\bbeta_1-\bbeta_{10})
+
\bC_{12}^n\hat\bbeta_2
-
\frac1n\bX_{n,1}^\top\boldsymbol\varepsilon_n
+
\lambda_n
\nabla_{\bbeta_1}P_m(\hat\bbeta)
=
\bzero.
$$
Multiplying by $\sqrt n$,
and since
$
\sqrt n\,\hat\bbeta_2=O_p(1)
$
and
$\bC_{12}^n\to\bC_{12}=\bzero$, gives
$
\bC_{12}^n\sqrt n\,\hat\bbeta_2=o_p(1),
$
we now have
$$
\sqrt n(\hat\bbeta_1-\bbeta_{10})
=
(\bC_{11}^n)^{-1}\bW_{n,1}
+
o_p(1),
$$
and it follows that
$
\sqrt n(\hat\bbeta_1-\bbeta_{10})
\xrightarrow{d}
\mathcal N_q
\left(
\bzero, \,
\sigma^2\bC_{11}^{-1}
\right).
$
Since
$
\binom{q-1}{m}=0
$ for $m > q$,
this is exactly the stated limiting distribution with
$\bd_{10}=\bzero$.
\qed

\section{Proof of Theorem \ref{thm:bss}}
Let $\bbeta^*=\hat\bbeta_m^{\mathrm{BSS}}$. Since $P_m(\beta^*) = 0$, global optimality of $\hat\bbeta_m(\lambda)$ gives
$$
\frac{1}{2n}\|\by_n - \bX_n\hat\bbeta_m(\lambda)\|_2^2
+\lambda P_m(\hat\bbeta_m(\lambda))
\le \frac{1}{2n}\|\by_n-\bX_n\bbeta^*\|_2^2.
$$
Therefore $P_m(\hat\bbeta_m(\lambda)) \to 0$ as $\lambda\to\infty$. Under the standing full-column-rank assumption, the least-squares term is coercive, and thus the global minimizers $\hat\bbeta_\lambda$ are uniformly bounded over $\lambda$. Let $\tilde\bbeta$ be any accumulation point along $\lambda_k\to\infty$ (as $k \to \infty$). Continuity of $P_m$ gives $P_m(\tilde\bbeta)=0$, hence $\|\tilde\bbeta\|_0\le m$.

For every $\bbeta \in \{\bbeta:\|\bbeta\|_0\le m\}$, we have $P_m(\bbeta) = 0$ and thus
$$
\frac{1}{2n}\|\by_n-\bX_n\hat\bbeta_m(\lambda_k)\|_2^2
\le
R_{\lambda_k,m}(\hat\bbeta_m(\lambda_k))
\le
\frac{1}{2n}\|\by_n-\bX_n\bbeta\|_2^2.
$$
Letting $k\to\infty$ yields
$$
\frac{1}{2n}\|\by_n-\bX_n\tilde\bbeta\|_2^2
\le
\frac{1}{2n}\|\by_n-\bX_n\bbeta\|_2^2
\qquad\text{for every }\bbeta \in \{\bbeta:\|\bbeta\|_0\le m\}.
$$
Thus $\tilde\bbeta$ is a best-subset minimizer. Uniqueness implies that every accumulation point is the same, and hence the whole path converges.\qed

\section{Best Subset Selection via Fridge}  \label{sec:bss_via_fridge}
\citet{zhu2020polynomial} introduced the special information criterion (SIC) to recover the true model size. Given an active set $\mathcal{A}$, SIC for the Fridge is formally defined as
\begin{align}
\sic_n(m) = n \log \left\{ (2n)^{-1} \|\by_n - \bX_n \hat\bbeta_{m}(\infty)\|_2^2 \right\} + m\log(p)\log\log(n).   \label{eq:fridge_sic}
\end{align}
Here, $\hat\bbeta_m(\infty)$ denotes any accumulation point of global $m$-Fridge minimizers as $\lambda\to\infty$, with $n$ fixed.

Let $\bX_{n,S}$ denote the submatrix formed by the columns indexed by $S$ and let
$\bbeta_{0,\mathcal A}$ be the nonzero coefficient vector on the true support $\mathcal A$.
Write $\gamma_n = \log(p_n)\log\log(n)$
and let $m_{max}=m_{max,n}$ satisfy $q\le m_{max}$ and ${m_{max}\gamma_n}/{n} = o(1).$
For example, if $\gamma_n=o(n)$, take
$m_{max}=\min\{p_n,\lfloor\sqrt{n/\gamma_n}\rfloor\}$.
Since $p_n\to\infty$ and $q$ is fixed, $m_{max}\to\infty$ and hence $m_{max}\ge q$ eventually, while
${m_{max}\gamma_n}/{n} \le \sqrt{{\gamma_n}/{n}} \to 0$.

We assume throughout this subsection that $p = p_n\to\infty$, and that $\mathcal A = \operatorname{supp}(\bbeta_0)$ has fixed size $q\ge1$. Assume that $\bX_n$ is deterministic and that $\by_n = \bX_n\bbeta_0+\boldsymbol\varepsilon_n$.
Assume the errors are independent, mean zero, have variance $\sigma^2>0$, and satisfy, for some constant $K<\infty$,
$\pr(|\varepsilon_i| > t) \le 2\exp(-t^2/K^2)$ for all $t>0$. We impose the following conditions:
\begin{enumerate}
	\item[(C1)] $m_{max}\gamma_n/n=o(1)$.
	\item[(C2)] There is a constant $M>0$, sufficiently large depending only on
	$c_{min}$, $\sigma^2$, and $K$, such that $\beta_{min,n}^2:=\min_{j\in\mathcal A}\beta_{0j}^2
	\ge M{\gamma_n}/n.$
	
	\item[(C3)] There exist constants $0<c_{min}\le c_{max}<\infty$ such that, for every
	$T\subset\{1,\ldots,p_n\}$ with $1\le |T|\le m_{max}+q$,
	$$
	c_{min}\le \lambda_{min}\!\left(n^{-1}\bX_{n,T}^\top\bX_{n,T}\right)
	\le \lambda_{max}\!\left(n^{-1}\bX_{n,T}^\top\bX_{n,T}\right)\le c_{max},
	$$
    where \(X_{n,T}\) denote the submatrix of \(X_n\) consisting of the columns indexed by \(T\).
\end{enumerate}

Under (C3) and finite positive penalty weights, accumulation points of $m$-Fridge as $\lambda\to\infty$ exist and solve the size-$m$ best-subset problem. We also note that penalized minimizers remain bounded, and that all such points give the same residual sum of squares for $\sic_n(m)$.

For each subset $S$ with $|S|\le m_{max}$, let $\widehat\beta_S^{\rm OLS} = \argmin_{b\in\mathbb R^{|S|}}\|\by_n-\bX_{n,S}b\|_2^2$, $\widehat\sigma_S^2 =\frac{1}{2n}\|\by_n-\bX_{n,S}\widehat\beta_S^{\rm OLS}\|_2^2,$ and define $\operatorname{SIC}_n(S)=n\log(\widehat\sigma_S^2)+|S|\gamma_n$ following \cite{zhu2020polynomial}.
Use the convention $\log 0=-\infty$ and a fixed rule to break ties.
Let $\widehat S_n\in\arg\min_{|S|\le m_{max}}\operatorname{SIC}_n(S)$ and $\widehat q_n=|\widehat S_n|$.
This subset formulation coincides with the version of selecting the model size from the
extreme-regularization ($\lambda=\infty$) Fridge fits \eqref{eq:fridge_sic}. It also avoids ambiguity when a
best-fit model under a target size $m$ happens to use fewer than $m$ nonzero coefficients.

In the following theorem, we show that the Fridge with target model size chosen via SIC yields the best subset selection property in high-dimensional setting.

\begin{theorem} \label{thm:best_m_adapt}
	Under (C1)--(C3), there exist constants $\alpha>0$, $C<\infty$, and $n_0$ such that, for all $n\ge n_0$,
$$
\pr\!\left(\widehat S_n=\mathcal A\right)
\ge 1-Cp_n^{-\alpha}.
$$
Consequently,
$$
\pr(\widehat q_n=q)\ge 1-Cp_n^{-\alpha}.
$$
\end{theorem}
The role of $\lambda\to\infty$ is only to identify the corresponding Fridge estimator with the best-subset least-squares problem used in the definition of $\operatorname{SIC}_n(S)$. No additional limit in $\lambda$ is taken in this statement.

\begin{proof}
Let $\Pi_S$ denote the orthogonal projection onto
$\operatorname{col}(\bX_{n,S})$, with $\Pi_S^\perp=I-\Pi_S$ and
$\Pi_{\varnothing}=0$. Write residual sums of squares
$$
R_S:=2n\widehat\sigma_S^2=\|\Pi_S^\perp\by_n\|_2^2,
\qquad
R_{\mathcal A}=\|\Pi_{\mathcal A}^\perp\by_n\|_2^2
=\boldsymbol\varepsilon_n^\top\Pi_{\mathcal A}^\perp\boldsymbol\varepsilon_n.
$$
All candidate models below satisfy $|S|\le m_{max}$. Let
$\mathcal E_{\rm over}$ and $\mathcal E_{\rm miss}$ be the events that some
$S\supsetneq\mathcal A$ (overfitting) or some $S\not\supseteq\mathcal A$ (misspecification), respectively,
satisfies $\operatorname{SIC}_n(S)\le\operatorname{SIC}_n(\mathcal A)$.
It suffices to bound the probabilities of these two events.

We first control the true-model residual sum of squares. Set
$c_R=\sigma^2/2$, $C_R=2\sigma^2$, and
$
\mathcal G_n=\{c_Rn\le R_{\mathcal A}\le C_Rn\}.
$
The projection $\Pi_{\mathcal A}^\perp$ has rank $n-q$, operator norm one, squared Frobenius norm $n-q$, and its quadratic form has expectation
$(n-q)\sigma^2$. Since $q$ is fixed, the Hanson-Wright inequality
\citep{rudelson2013hanson} gives, for all large $n$,
$$
\pr(\mathcal G_n^c)\le C\exp(-cn).
$$
Condition (C1) and $m_{max}\ge q\ge1$ imply $\gamma_n=o(n)$ and hence
$\log p_n=o(n)$. Thus this exponential error is $O(p_n^{-\alpha})$
for any fixed $\alpha>0$.

\medskip
\noindent\textbf{Case 1: $S\supsetneq\mathcal A$.}

Write $|S|=q+k$, $k\ge1$, and put
$H_S := \Pi_S-\Pi_{\mathcal A}$, $T_S := \boldsymbol\varepsilon_n^\top H_S\boldsymbol\varepsilon_n$.
Then, $H_S$ is an orthogonal projection of rank $k$ and
$R_{\mathcal A}-R_S=T_S$. On $\mathcal G_n$, if
$T_S\le(c_R/4)k\gamma_n$, then
$$
0\le\frac{T_S}{R_{\mathcal A}}
\le\frac{k\gamma_n}{4n}=o(1)
$$
uniformly over these models by (C1). For all sufficiently large $n$,
the inequality $-\log(1-x)\le2x$ for $0\le x\le1/2$ therefore gives
\begin{align*}
\operatorname{SIC}_n(\mathcal A)-\operatorname{SIC}_n(S)
&=n\log\left(1-\frac{T_S}{R_{\mathcal A}}\right)^{-1}-k\gamma_n\\
&\le\frac{2}{c_R}T_S-k\gamma_n
\le-\frac12k\gamma_n<0.
\end{align*}
Since $\mathbb E T_S=\sigma^2k$, $\|H_S\|_{\rm op}=1$, and
$\|H_S\|_F^2=k$, Hanson--Wright \citep[Theorem 1.1]{rudelson2013hanson} gives a constant $c_*>0$ such that
$$
\pr\{T_S>(c_R/4)k\gamma_n\}\le2\exp(-c_*k\gamma_n)
$$
for all sufficiently large $n$, uniformly over $S$. Here
$\gamma_n\to\infty$ ensures that the threshold exceeds the expectation.
There are at most $p_n^k$ such models of size $q+k$, so
\begin{align*}
\pr(\mathcal E_{\rm over})
&\le \pr(\mathcal G_n^c)
+ 2\sum_{k=1}^{m_{max}-q}\exp\{k\log p_n-c_*k\gamma_n\}\\
&\le C\exp(-cn)
+ 2\sum_{k=1}^{\infty}p_n^{-k\{c_*\log\log n-1\}}
= O(p_n^{-\alpha})
\end{align*}
for some $\alpha>0$.

\medskip
\noindent\textbf{Case 2: $S\not\supseteq\mathcal A$.}

The model $S$ omits at least one true variable and may also include noise variables. Write
$$
U=\mathcal A\setminus S,\qquad V=S\setminus\mathcal A,
\qquad k=|U|\ge1,\qquad v=|V|,\qquad r=k+v.
$$
Thus $k$ counts omitted true variables, $v$ counts included noise variables, and $|S|-q=v-k$.
We first bound the signal lost by omitting $U$, then control the noise terms simultaneously over all candidate models.

The signal that cannot be explained by $S$ is $\Delta_S=\|a_S\|_2^2$ where $a_S=\Pi_S^\perp\bX_{n,U}\bbeta_{0,U}$.
Projection onto $\operatorname{col}(\bX_{n,S})$ gives the least-squares approximation, so $\Delta_S=\min_b\|\bX_{n,U}\bbeta_{0,U}-\bX_{n,S}b\|_2^2$.
For every $b$, the vector inside this norm is a linear combination of the columns indexed by $U\cup S=S\cup\mathcal A$, with coefficient vector $(\bbeta_{0,U}^\top,-b^\top)^\top$.
Since $|S\cup\mathcal A|\le m_{max}+q$, the lower eigenvalue bound in (C3) gives
$$
\|\bX_{n,U}\bbeta_{0,U}-\bX_{n,S}b\|_2^2
\ge nc_{min}\bigl(\|\bbeta_{0,U}\|_2^2+\|b\|_2^2\bigr)
\ge nc_{min}\|\bbeta_{0,U}\|_2^2.
$$
Taking the minimum over $b$ and using (C2), we obtain
\begin{equation}
\Delta_S\ge nc_{min}k\beta_{min,n}^2
\ge c_{min}Mk\gamma_n.
\label{eq:Delta_underfit}
\end{equation}

To identify the noise terms, note that
$\Pi_S^\perp\by_n=a_S+\Pi_S^\perp\boldsymbol\varepsilon_n$ and
$\Pi_S^\perp a_S=a_S$. Expanding the squared norm and subtracting $R_{\mathcal A}$ gives
\begin{equation}
R_S-R_{\mathcal A}
=\Delta_S+2\boldsymbol\varepsilon_n^\top a_S
+\boldsymbol\varepsilon_n^\top B_S\boldsymbol\varepsilon_n,
~~ \text{where } B_S=\Pi_{\mathcal A}-\Pi_S.
\label{eq:misspecified_rss_decomposition}
\end{equation}
The first term is the loss from the omitted signals. The remaining two terms describe how noise changes the residual comparison.

For a fixed $S$, the vector $a_S$ is deterministic and
$\boldsymbol\varepsilon_n^\top a_S=\sum_{i=1}^n(a_S)_i\varepsilon_i$ is a weighted sum of independent, centered sub-Gaussian errors.
The general Hoeffding inequality \citep[Theorem 2.6.3]{vershynin2018high} therefore gives
$$
\pr\bigl(|\boldsymbol\varepsilon_n^\top a_S|>t\bigr)
\le2\exp\left(-\frac{ct^2}{\Delta_S}\right),\qquad t>0,
$$
where $c>0$ depends only on $K$.
Fix $L>2$ and choose $A_L$ sufficiently large. Substituting
$t=A_L\sqrt{\Delta_S r\log p_n}$ makes this probability at most $2p_n^{-Lr}$.
Moreover,
$$
2A_L\sqrt{\Delta_S r\log p_n}
\le\frac14\Delta_S+4A_L^2r\log p_n.
$$
Thus, with $C_4=4A_L^2$, the inequality
\begin{equation}
2|\boldsymbol\varepsilon_n^\top a_S|
\le\frac14\Delta_S+C_4r\log p_n
\label{eq:cross_uniform}
\end{equation}
fails for this $S$ with probability at most $2p_n^{-Lr}$.

For the quadratic noise term, we first state the properties of $B_S$ needed for concentration.
For every unit vector $h$,
$
h^\top B_Sh=\|\Pi_{\mathcal A}h\|_2^2-\|\Pi_Sh\|_2^2\in[-1,1].
$
Because $B_S$ is symmetric, all its eigenvalues lie in $[-1,1]$, so $\|B_S\|_{2} \le 1$.
Both projections act as the identity on $\operatorname{col}(\bX_{n,S\cap\mathcal A})$, so their difference vanishes there.
By symmetry, the range of $B_S$ is orthogonal to this common subspace. It also lies in $\operatorname{col}(\bX_{n,S\cup\mathcal A})$.
(C3) makes the columns in the union linearly independent. These two subspaces have dimensions $q-k$ and $q+v$, respectively, and hence
$
\operatorname{rank}(B_S)\le(q+v)-(q-k)=r.
$
Also,
$
\|B_S\|_F^2=\sum_i\lambda_i(B_S)^2
\le\operatorname{rank}(B_S)\le r,
$ where $\lambda_i(\cdot)$ denotes the $i$th eigenvalue.

The errors have covariance $\sigma^2 I_n$ and mean zero. Therefore,
\begin{align*}
\E\bigl(\boldsymbol\varepsilon_n^\top B_S\boldsymbol\varepsilon_n\bigr)
=\sigma^2\operatorname{tr}(B_S)
=\sigma^2(q-|S|)=\sigma^2(k-v).
\end{align*}
Using the operator and Frobenius norm bounds above, the Hanson--Wright inequality
\citep[Theorem 1.1]{rudelson2013hanson} gives
$$
\pr\left(\left|\boldsymbol\varepsilon_n^\top B_S\boldsymbol\varepsilon_n
-\sigma^2(k-v)\right|>t\right)
\le2\exp\left\{-c\min\left(\frac{t^2}{r},t\right)\right\},\qquad t>0,
$$
where $c>0$ depends only on $K$.
Choose $t$ to be a sufficiently large constant multiple of $r\log p_n$.
Then for a sufficiently large $C_5$,
\begin{equation}
|\boldsymbol\varepsilon_n^\top B_S\boldsymbol\varepsilon_n|
\le C_5r\log p_n
\label{eq:quad_uniform}
\end{equation}
fails for this $S$ with probability at most $2p_n^{-Lr}$, for all sufficiently large $n$.

We now make these bounds simultaneous. Define $\mathcal E_n$ to be the event that both \eqref{eq:cross_uniform} and \eqref{eq:quad_uniform} hold for every $S\not\supseteq\mathcal A$ with $|S|\le m_{max}$.
For given $k,v$, there are
$\binom qk\binom{p_n-q}{v}\le p_n^{k+v}$ possible models.
For each such model, the combined failure probability is at most $4p_n^{-L(k+v)}$.
The union bound yields
\begin{align*}
\pr(\mathcal E_n^c)
\le4\sum_{k=1}^{q}\sum_{v=0}^{m_{max}}p_n^{-(L-1)(k+v)}
\le\frac{4p_n^{-(L-1)}}{\{1-p_n^{-(L-1)}\}^2}
=O\bigl(p_n^{-(L-1)}\bigr).
\end{align*}

On $\mathcal E_n$, substitute the two noise bounds into
\eqref{eq:misspecified_rss_decomposition}. With $C_7=C_4+C_5$, this gives
\begin{align}
R_S-R_{\mathcal A}
&\ge\frac34\Delta_S-C_7r\log p_n,
\label{eq:D_underfit}\\
|R_S-R_{\mathcal A}|
&\le\frac54\Delta_S+C_7r\log p_n
\le C_6(\Delta_S+r\log p_n),
\label{eq:D_two_sided}
\end{align}
where $C_6=\max(5/4,C_7)$.
These inequalities hold for all misspecified models at once.

Considering only $\mathcal E_n\cap\mathcal G_n$, the preceding bounds hold and
$c_Rn\le R_{\mathcal A}\le C_Rn$.
Note that we have
$\pr\{(\mathcal E_n\cap\mathcal G_n)^c\} \le \pr(\mathcal E_n^c)+\pr(\mathcal G_n^c)=O(p_n^{-\alpha})$
for some $\alpha>0$, decreased if necessary to also apply in Case 1.
We will show that every misspecified model has larger SIC on this event.
The two cases below distinguish whether the loss from the omitted signals or the number of added noise variables provides the decisive comparison.

\medskip
\noindent\textbf{Case 2a: $C_7r\log p_n\le(3/8)\Delta_S$.}

Then $R_S-R_{\mathcal A}\ge(3/8)\Delta_S>0$. Since $R_{\mathcal A}\le C_Rn$ and
$\log(1+x)\ge(\log2)\min(x,1)$ for $x\ge0$, there are constants
$c_8,c_9>0$, depending only on $\sigma^2$, such that
\begin{align}
n\log\left(\frac{R_S}{R_{\mathcal A}}\right)
=n\log\left(1+\frac{R_S-R_{\mathcal A}}{R_{\mathcal A}}\right)
\ge c_8\min(R_S-R_{\mathcal A},n)
\ge c_9\min(\Delta_S,n).
\label{eq:underfit_signal_log}
\end{align}
It follows that
$\operatorname{SIC}_n(S)-\operatorname{SIC}_n(\mathcal A) \ge c_9\min(\Delta_S,n)+(v-k)\gamma_n$.
If $v\ge k$, the right-hand side is strictly positive.
If $v<k$ and $\Delta_S\le n$, \eqref{eq:Delta_underfit} gives
$$
\operatorname{SIC}_n(S)-\operatorname{SIC}_n(\mathcal A)
\ge(c_9c_{min}M-1)k\gamma_n>0
$$
when $M$ is sufficiently large. If $v<k$ and $\Delta_S>n$, the same
difference is at least $c_9n-k\gamma_n>0$ for all sufficiently large $n$,
since $k\le q$ is fixed and $\gamma_n=o(n)$.

\medskip
\noindent\textbf{Case 2b: $C_7r\log p_n>(3/8)\Delta_S$.}

Combining this condition with \eqref{eq:Delta_underfit} gives
\begin{equation}
r > \frac{3c_{min}M}{8C_7}k\log\log n.
\end{equation}
Thus $r/k\to\infty$ uniformly over models in this case. For all
sufficiently large $n$, $r\ge4k$ and hence
\begin{equation}
v-k=r-2k\ge r/2.
\label{eq:vk_lower}
\end{equation}
For every $S$ in Case 2b, its defining inequality gives
$\Delta_S<(8/3)C_7r\log p_n$. Combining this with \eqref{eq:D_two_sided} yields
$|R_S-R_{\mathcal A}|\le C_{12}r\log p_n$,
where $C_{12}=C_6(1+(8/3)C_7)$. Since $r\le m_{max}+q$ and
$R_{\mathcal A}\ge c_Rn$ on $\mathcal E_n\cap\mathcal G_n$, (C1) gives the following
bound uniformly over all models in Case 2b:
$$
\left|\frac{R_S-R_{\mathcal A}}{R_{\mathcal A}}\right|
\le\frac{C_{12}(m_{max}+q)\log p_n}{c_Rn}=o(1).
$$
For all sufficiently large $n$, the left-hand side is at most $1/2$. Using $\log(1+x)\ge-2|x|$ for $|x|\le1/2$, we obtain
\begin{equation*}
n\log\left(\frac{R_S}{R_{\mathcal A}}\right)
\ge -\frac{2n}{R_{\mathcal A}}|R_S-R_{\mathcal A}|
\ge -C_{13}r\log p_n.
\end{equation*}
Together with \eqref{eq:vk_lower}, this gives
\begin{align*}
\operatorname{SIC}_n(S)-\operatorname{SIC}_n(\mathcal A)
&\ge-C_{13}r\log p_n+(v-k)\gamma_n\\
&\ge r\log p_n\left\{\frac12\log\log n-C_{13}\right\}>0
\end{align*}
for all sufficiently large $n$.

Thus every misspecified model has strictly larger SIC than $\mathcal A$
on $\mathcal E_n\cap\mathcal G_n$, so
$\pr(\mathcal E_{\rm miss})\le\pr\{(\mathcal E_n\cap\mathcal G_n)^c\}=O(p_n^{-\alpha})$.
Finally,
$$
\pr(\widehat S_n\ne\mathcal A)
\le \pr(\mathcal E_{\rm over}) + \pr(\mathcal E_{\rm miss})
\le Cp_n^{-\alpha}.
$$
Since $\widehat S_n=\mathcal A$ implies $\widehat q_n=q$, the model-size
conclusion follows as well.
\end{proof}

\section{Property of Fridge When $g_j(b)=|b|$}  \label{A:stdfridge}
\subsection{Limiting Distribution} \label{A:limiting}

First, we discuss how the limiting distribution changes with the choice of $g_j(b)=|b|$. 

\begin{theorem}[Limiting distribution]  
	Let $p$ be fixed. Assume that $\bC_{12}=\bzero$ and ${\sqrt{n}}{\lambda_n} \to \lambda_0 \ge 0$ as $n \to \infty$. 
	For a fixed $m$, let $\hat{\bbeta} = (\hat{\bbeta}_1^\top ~ \hat{\bbeta}_2^\top)^\top$ be a global minimizer of $R_n(\bbeta)=(2n)^{-1}\|\by_n-\bX_n\bbeta\|_2^2+\lambda_nP_m(|\bbeta|)$. We have
	$$\sqrt{n}(\hat{\bbeta}_1 - \boldsymbol{\beta}_{10}) \xrightarrow{d} \mathcal{N}_q\left(-{\lambda_0} \bC_{11}^{-1} \bd_{10}, \; \sigma^2 \bC_{11}^{-1}\right),$$
	where the $j$-th element of $\bd_{10}$ is given as
	$$
	\sign(\beta_{0j}) \sum_{\substack{S \subseteq \mathcal{A} \setminus \{j\} \\ |S| = m}} \prod_{k \in S} |\beta_{0k}|.
	$$
\end{theorem}

\begin{corollary}
	Under the assumptions of the theorem,
	\begin{enumerate}
		\item[(i)] If $m \ge q$, then $\bd_{10} = \bzero$, and the active set coefficient estimators are asymptotically unbiased.
		\item[(ii)] If $m \le q$ and $\lambda_0 > 0$, then $\pr(\hat{\bbeta}_2 = \bzero)$ converges to a strictly positive value.
	\end{enumerate}
\end{corollary}

\begin{proof}
Most of the arguments are the same as in Theorem \ref{thm:limiting}. Only the penalty contribution changes: the active derivative is now the unweighted $\bd_{10}$, and, for $m\le q$, the inactive contribution has a finite $\ell_1$ limit instead of forcing all inactive coordinates to zero.

By consistency, local Lipschitz continuity of the unweighted penalty and global optimality, with probability tending to one,
$$
\|\hat\bbeta-\bbeta_0\|_2
\le\frac{2}{c}\left(\|n^{-1}\bX_n^\top\boldsymbol\varepsilon_n\|_2+\lambda_nL\right)
=O_p(n^{-1/2}),
$$
where $L$ is a local Lipschitz constant and $c>0$ bounds the smallest eigenvalue of $\bC_n$ below. The restricted estimator $(\tilde\bbeta_1,0)$ satisfies the same rate.

Write $\bW_n=n^{-1/2}\bX_n^\top\boldsymbol\varepsilon_n\xrightarrow{d}\bW\sim\mathcal N_p(\bzero,\sigma^2\bC)$ and $D_m=P_{m-1}(|\bbeta_{10}|)$, taking $D_m=0$ for $m>q$ and $P_{-1}\equiv1$. Uniformly for bounded $\bu=(\bu_1^\top,\bu_2^\top)^\top$,
$$
\sqrt n\left\{P_m(|\bbeta_0+\bu/\sqrt n|)-P_m(|\bbeta_0|)\right\}
= \bd_{10}^\top\bu_1+D_m\|\bu_2\|_1+o(1).
$$
Products with zero or one inactive factor give the two displayed terms; products with at least two are $O(n^{-1})$. Combining this expansion with the quadratic loss expansion yields the limiting local objective
$$
V(\bu)=\tfrac12\bu^\top\bC\bu-\bu^\top\bW
+\lambda_0\{\bd_{10}^\top\bu_1+D_m\|\bu_2\|_1\}.
$$
The expansion holds uniformly on every fixed bounded set, and $\sqrt n(\hat\bbeta-\bbeta_0)=O_p(1)$. Since $C$ is positive definite and $\lambda_0D_m \ge 0$, $V$ is strongly convex and coercive, and hence has a unique minimizer. Applying the argmax theorem to the negatives of the local objectives \citep[Section 5.9]{vandervaart1998asymptotic} therefore gives $\sqrt n(\hat\bbeta-\bbeta_0)\xrightarrow{d}\arg\min_{\bu}V(\bu)$. Since $\bC_{12}=0$, its active component is $\bC_{11}^{-1}(\bW_1-\lambda_0\bd_{10})$, proving Part (i) of the theorem. For $m>q$, both penalty terms vanish.

For Part (ii), the threshold now has a finite limit. On a fixed bounded convex neighborhood of $\bbeta_0$, the unweighted penalty Hessians, where defined, have operator norm bounded by a constant $K$. Since $\lambda_{\min}(\bC_n)\ge c>0$ and $\lambda_n\to0$, the smooth Hessian of $R_n$ is bounded below by $(c-\lambda_nK)I\succeq(c/2)I$ for all sufficiently large $n$. The nonnegative slope jumps at zero coordinates preserve this bound across orthants, similar to the proof of Lemma \ref{lem:local_convex}. Thus $R_n$ is strongly convex on this neighborhood. Together with consistency, this makes $\hat\bbeta_2=0$ equivalent, with probability tending to one, to
$$
\left\|n^{-1/2}\bX_{n,2}^\top(\by_n-\bX_{n,1}\tilde\bbeta_1)\right\|_\infty
\le\sqrt n\lambda_nP_{m-1}(|\tilde\bbeta_1|).
$$
The score vector on the left is $\bW_{n,2}-\bC_{21}^n\sqrt n(\tilde\bbeta_1-\bbeta_{10})=\bW_{n,2}+o_p(1)$, and the threshold tends to $\lambda_0D_m$. The Gaussian limit assigns zero probability to the boundary, so for $m\le q,\ \lambda_0>0$,
$$
\pr(\hat\bbeta_2=\bzero) \to
\pr(\|\bW_2\|_\infty\le\lambda_0D_m)>0.
$$
\end{proof}

For $m\le q$, if $\bC_{22}$ is diagonal, the individual inactive-coordinate KKT conditions yield
\begin{align}
	\pr(\hat{\bbeta}_{2j} = 0) \to 2\Phi\left( \frac{\lambda_0 D_m}{\sigma \sqrt{(\bC_{22})_{jj}}} \right) - 1,   \label{eq:sparsity}
\end{align}
Here $D_m=\sum_{S\subseteq\mathcal A,\,|S|=m}\prod_{k\in S}|\beta_{0k}|$. If $\bC_{22}=\bI$ and $|\beta_{0k}|=c>0$ for all $k\in\mathcal A$, then $D_m={q\choose m}c^m$ and the limiting probability of excluding all $p-q$ inactive coefficients is $[2\Phi(\lambda_0D_m/\sigma)-1]^{p-q}$. This joint probability can be close to one in some cases, for example, with $q=m=3$, $p=13$, $\lambda_0=\sigma=1$, and $c=1.5$, the probability is $[2\Phi(3.375)-1]^{10} \approx 0.993$.

\subsection{Irrepresentable Condition}

The next theorem states that Fridge is somewhat robust to predictor collinearity.

\begin{theorem}  \label{thm:irrep}
Let $m<q$, $P_{-1}\equiv1$, and suppose the active coefficients are bounded away from zero. Suppose $\lambda_n\to0$ and $\sqrt n\lambda_n\to\infty$. If, for some $\eta>0$,
$$
\left\|\bC_{21}^n(\bC_{11}^{n})^{-1} \nabla_{\mathcal A}P_m(\bbeta_0) \right\|_\infty\le(1-\eta) P_{m-1}(|\bbeta_{10}|),
$$
then the oracle active-sign solution satisfies the inactive KKT inequalities with probability tending to one. Under local strict convexity of the active block, there is consequently no false-positive coordinate in the corresponding local minimizer with probability tending to one.
\end{theorem}
If $|\beta_{0j}|=c$ for all $j\in\mathcal A$, then the condition reduces to
\begin{equation}
\left\|\bC_{21}^n(\bC_{11}^{n})^{-1}\sign(\bbeta_{10})\right\|_\infty
\le (1-\eta)\frac{q}{q-m}.
\label{eq:irrep}
\end{equation}
This corresponds with the Lasso's ``irrepresentable condition'' \citep{zhao2006model}. Lasso instead requires the right-hand side of \eqref{eq:irrep} to be strictly less than $1-\eta$. This shows that whenever the target model size is chosen smaller than $q$, the model is somewhat robust against the correlation between signal and noise variables. If $m=q-1$, the right-hand side of \eqref{eq:irrep} is the largest with $(1-\eta)q$.

\begin{proof}
For simplicity, let $d=\nabla_{\mathcal A}P_m(\bbeta_0)$ and $D = P_{m-1}(|\bbeta_{10}|)$. Let $\tilde\bbeta_1$ denote the minimizer of the objective restricted to the active coordinates with the inactive coordinates fixed at zero. Its active first-order condition implies
$$
\tilde\bbeta_1-\bbeta_{10}
=(\bC_{11}^n)^{-1}\left\{n^{-1}\bX_1^\top\boldsymbol{\varepsilon}_n-\lambda_n d\right\}+o_p(n^{-1/2}+\lambda_n).
$$
Consequently,
$$
n^{-1}\bX_2^\top(\by_n-\bX_1\tilde\bbeta_1)
=\lambda_n\bC_{21}^n(\bC_{11}^n)^{-1}d+O_p(n^{-1/2})+o_p(\lambda_n).
$$
Since $\sqrt n\lambda_n\to\infty$, the stochastic score is $o_p(\lambda_n)$. The inactive KKT condition at zero is
$$
\left|n^{-1}\bx_j^\top(\by_n-\bX_1\tilde\bbeta_1)\right| \le \lambda_nP_{m-1}(|\tilde\bbeta_1|)=\lambda_nD+o_p(\lambda_n).
$$
The strict margin $(1-\eta)D$ therefore makes all inactive KKT inequalities hold with probability tending to one. Local strict convexity on the active block then gives the stated local-minimizer conclusion.
\end{proof}

\subsection{Best Subset Selection}
The finite-sample extreme-regularization conclusion of Theorem \ref{thm:bss} remains valid for the case $g_j(b) = |b|$.

\section{Details About Numerical Studies in Section \ref{sec:sim}}  \label{A:sim_detail}

\subsection{Data Generating Process}
We generated synthetic datasets from a linear regression model, $y = X\beta + \epsilon$. The process is specifically designed to construct highly correlated feature blocks. 

We simulate $n = n_{train} + n_{test}$ observations with $p$ predictors. The rows of the design matrix $\bX \in \mathbb{R}^{N \times p}$ are drawn independently from a multivariate normal distribution $\mathcal{N}_p(0, \Sigma)$.
To induce multicollinearity, the covariance matrix $\Sigma$ is structured as an AR(1) process, where the correlation between the $i$-th and $j$-th predictors decays exponentially with their positional distance. The entries of $\Sigma$ are $\Sigma_{i,j} = \rho^{|i - j|},$ where $\rho \in [0, 1)$ controls the baseline correlation strength. After generation, the entire design matrix $\bX$ is column-standardized so that each predictor has an empirical mean of $0$ and a standard deviation of $1$.

The true coefficient vector $\bbeta_{0} \in \mathbb{R}^p$ is designed to be strictly sparse, containing exactly $q$ non-zero entries (the active set).
To control the degree of correlation among the active predictors, their indices are spaced out by a constant integer, step. The indices of the active variables are defined by the sequence $\{0, \text{step}, 2\times\text{step}, \dots, (q-1)\times\text{step}\}$. When $\text{step} = 1$, the active variables form a contiguous block $\{0, 1, \dots, q-1\}$ that are highly correlated due to the AR(1) structure of $X$.
For each index in the active set, the coefficient magnitude is fixed at exactly $1$, while the sign is randomly drawn with equal probability from $\{-1, 1\}$. That is, for $j$ in the active set:
$$\beta_j = c_j \times 1.0, \quad c_j \sim \text{Uniform}\{-1, 1\}.$$
All other coefficients not in the active set are set to $0$.

To strictly control the complexity of the task, we define the SNR based on the empirical variances of the generated data. The deterministic signal vector is calculated as $f(\bX) = \bX\bbeta_{0}$.
Given a target $\text{SNR}$, the required variance for the random error term, $\sigma^2$, is calculated analytically as:
$$\sigma^2 = \frac{\bbeta_{0}^\top\bX^\top\bX\bbeta_{0}/n}{\text{SNR}}.$$
The noise vector $\boldsymbol\epsilon \in \mathbb{R}^n$ is then generated from independent and identically distributed normal draws, $\boldsymbol\epsilon \sim \mathcal{N}_n(\bzero, \sigma^2\bI)$.

The final response vector $\by$ is computed by adding the noise to the true signal: $\by = \bX\bbeta_{0} + \boldsymbol\epsilon.$
The response $\by$ is then mean-centered. Finally, the $n$ observations are partitioned sequentially: the first $n_{train}$ observations are allocated to the training set $(\bX_{train}, \by_{train})$ for model fitting, and the remaining $n_{test}$ observations are reserved for the testing set $(\bX_{test}, \by_{test})$ to evaluate out-of-sample predictive performance.

\subsection{Evaluation Metrics} 
Following \citet{hastie2020best}, we considered the following evaluation metrics:
\begin{itemize}
	\item F-score: 
	\begin{align*}
		\text{F-score}
		= \left( \frac{\text{recall}^{-1} + \text{precision}^{-1}}{2} \right)^{-1}.
	\end{align*}
	A perfect score is $1$.
	\item Proportion of variance explained (PVE):
	\begin{align*}
		\text{PVE}(\hat\bbeta)
		&= 1 - \frac{(\hat\bbeta - \bbeta_0)^\top \Sigma (\hat\bbeta - \bbeta_0) + \sigma^2}{\bbeta_0^\top\Sigma\bbeta_0 + \sigma^2}.
	\end{align*}
	A perfect score is $\text{SNR}/(1+\text{SNR})$ and the null score is $0$. This metric can be negative if a model performs worse than the null model.
	\item Relative test error (RTE): this metric measures the expected test error relative to the Bayes error rate.
	\begin{align*}
		\text{RTE}(\hat\bbeta)
		&= \frac{(\hat\bbeta - \bbeta_0)^\top \Sigma (\hat\bbeta - \bbeta_0) + \sigma^2}{\sigma^2}.
	\end{align*}
	A perfect score is $1$ and the null score is SNR $+ 1$.
	\item Model size: the number of non-zero coefficient estimates.
\end{itemize}

\subsection{Simulation Results}
\begin{figure}[!htp]
\centering
  \includegraphics[width=\textwidth]{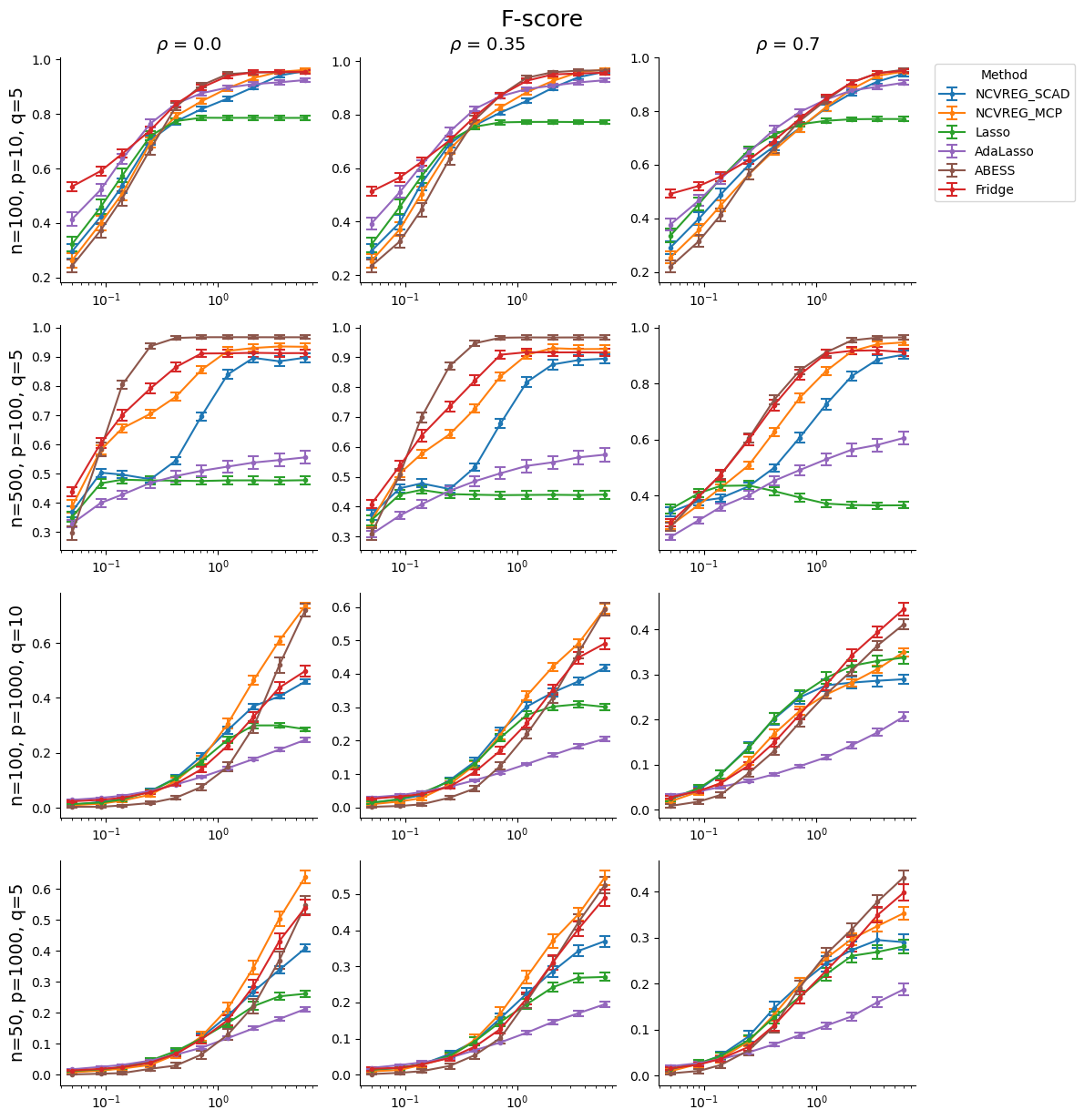}
  \caption[F-score in the setting with step $=1$.]{
    F-score (ordinate) versus SNR (abscissa) for Step $=1$. The ten SNR values range from $0.05$ to $6$ and are equally spaced on a $\log_{10}$ scale. Rows correspond to the indicated $(n,p,q)$ settings, and columns to predictor correlations $\rho=0$, $0.35$, and $0.7$. The ordinate is linear, with axis limits varying across panels. Points show means over 500 independent Monte Carlo replications; error bars indicate $\pm 2$ standard errors.
  }
  \label{fig:step1_fscore}
\end{figure}

\begin{figure}[!htp]
\centering
  \includegraphics[width=\textwidth]{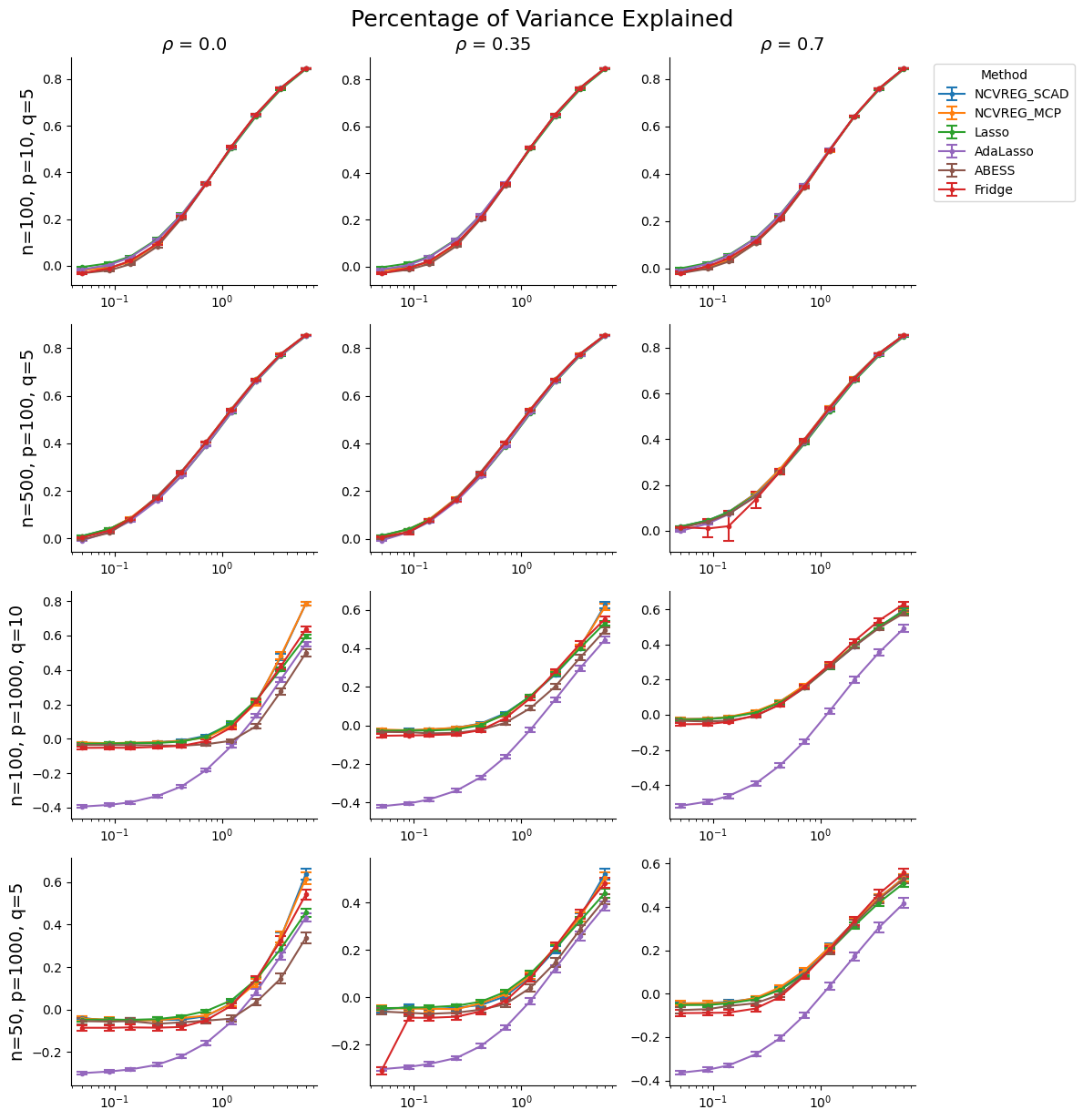}
  \caption[PVE in the setting with step $=1$.]{
    Proportion of variance explained (PVE) (ordinate) versus SNR (abscissa) for Step $=1$. The ten SNR values range from $0.05$ to $6$ and are equally spaced on a $\log_{10}$ scale. Rows correspond to the indicated $(n,p,q)$ settings, and columns to predictor correlations $\rho=0$, $0.35$, and $0.7$. The ordinate is linear, with axis limits varying across panels. Points show means over 500 independent Monte Carlo replications; error bars indicate $\pm 2$ standard errors. 
  }
  \label{fig:step1_pve}
\end{figure}

\begin{figure}[!htp]
\centering
  \includegraphics[width=\textwidth]{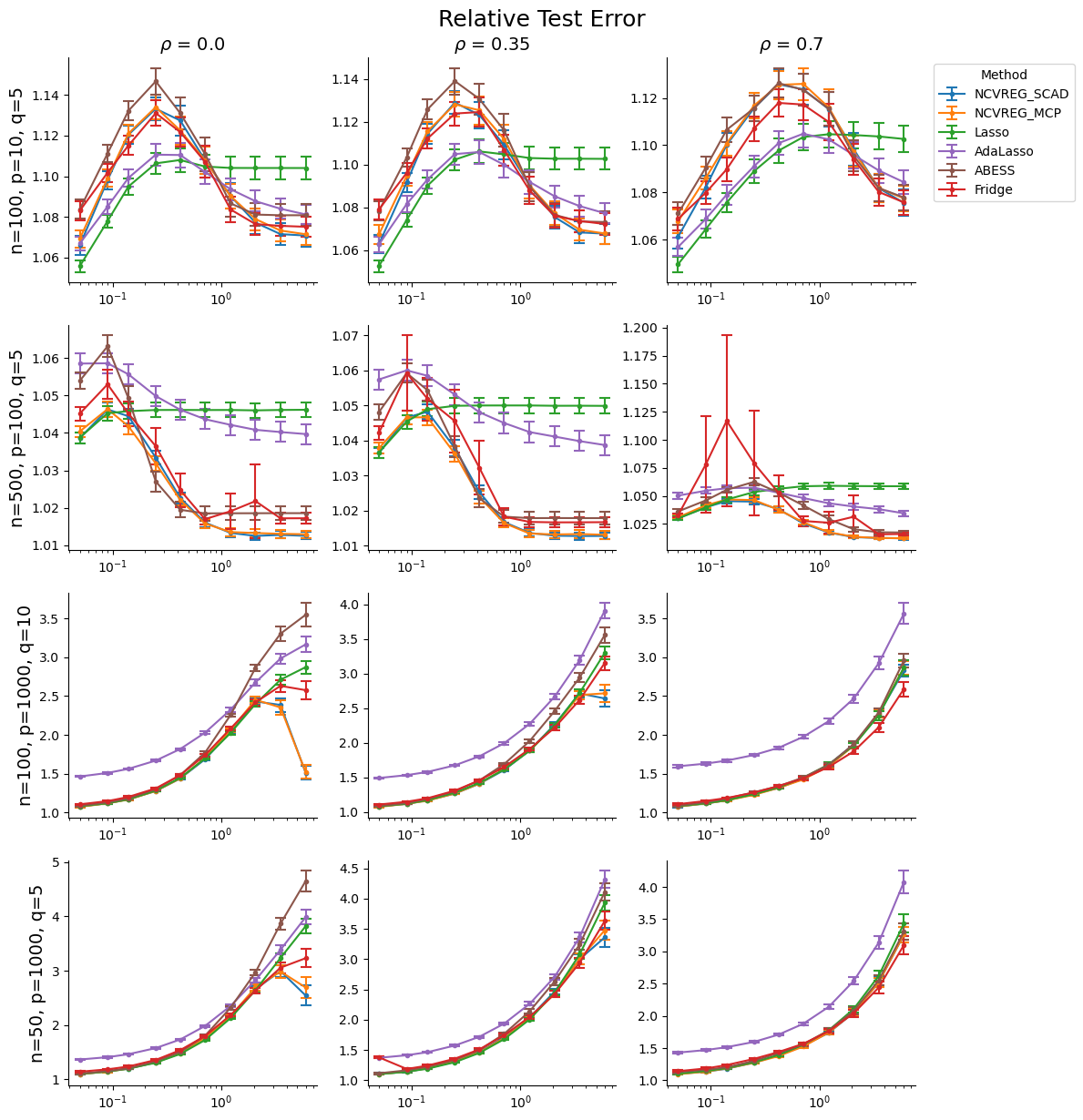}
  \caption[RTE in the setting with step $=1$.]{
    Relative test error (RTE) (ordinate) versus SNR (abscissa) for Step $=1$. The ten SNR values range from $0.05$ to $6$ and are equally spaced on a $\log_{10}$ scale. Rows correspond to the indicated $(n,p,q)$ settings, and columns to predictor correlations $\rho=0$, $0.35$, and $0.7$. The ordinate is linear, with axis limits varying across panels. Points show means over 500 independent Monte Carlo replications; error bars indicate $\pm 2$ standard errors.
  }
  \label{fig:step1_rte}
\end{figure}

\begin{figure}[!htp]
\centering
  \includegraphics[width=\textwidth]{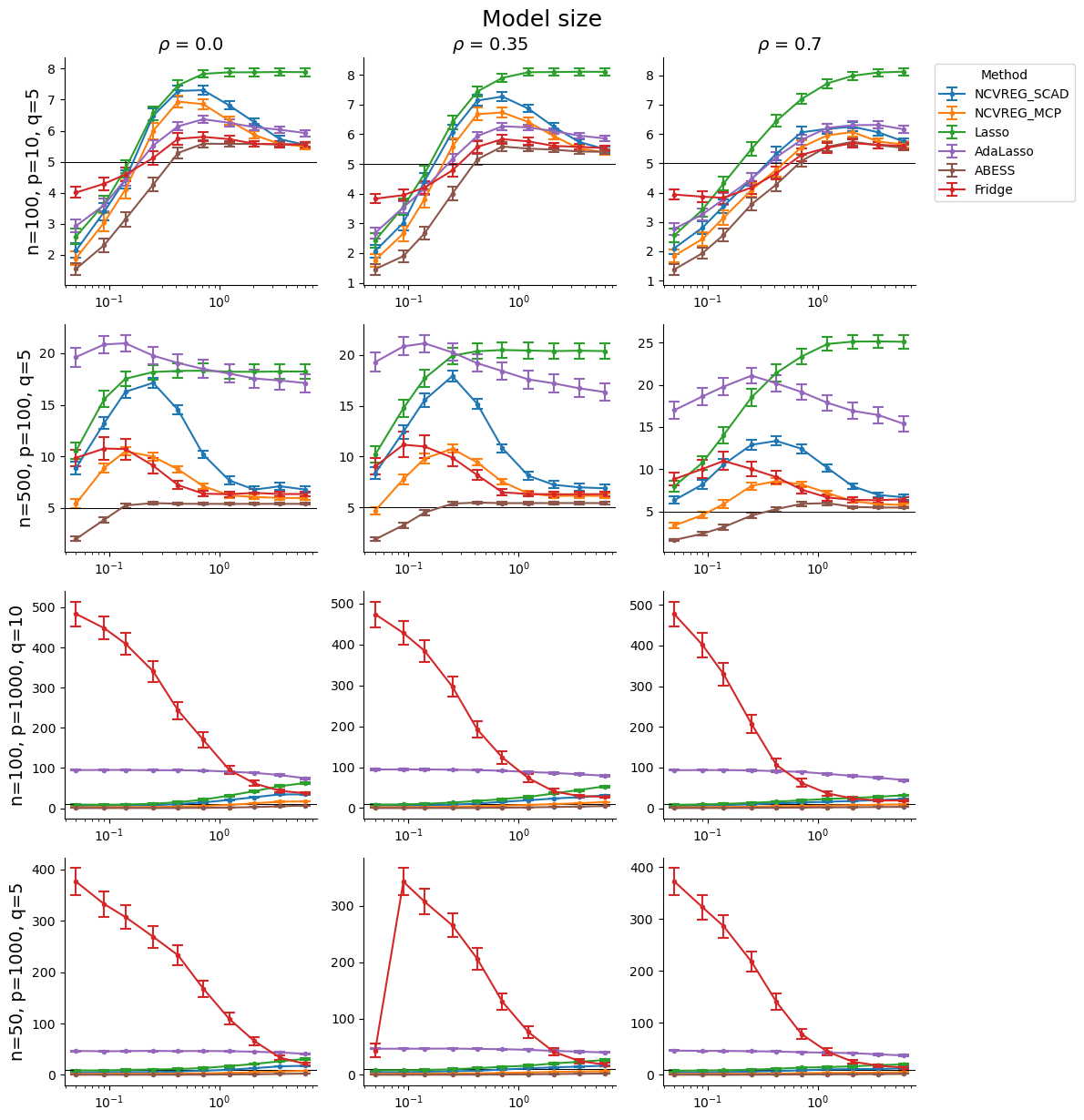}
  \caption[Model size in the setting with step $=1$.]{
    Selected model size (ordinate) versus SNR (abscissa) for Step $=1$. The ten SNR values range from $0.05$ to $6$ and are equally spaced on a $\log_{10}$ scale. Rows correspond to the indicated $(n,p,q)$ settings, and columns to predictor correlations $\rho=0$, $0.35$, and $0.7$. The ordinate is linear, with axis limits varying across panels. Points show means over 500 independent Monte Carlo replications; error bars indicate $\pm 2$ standard errors. The solid black line indicates the true model size $q$.
  }
  \label{fig:step1_modelsize}
\end{figure}

\begin{figure}[!htp]
\centering
  \includegraphics[width=\textwidth]{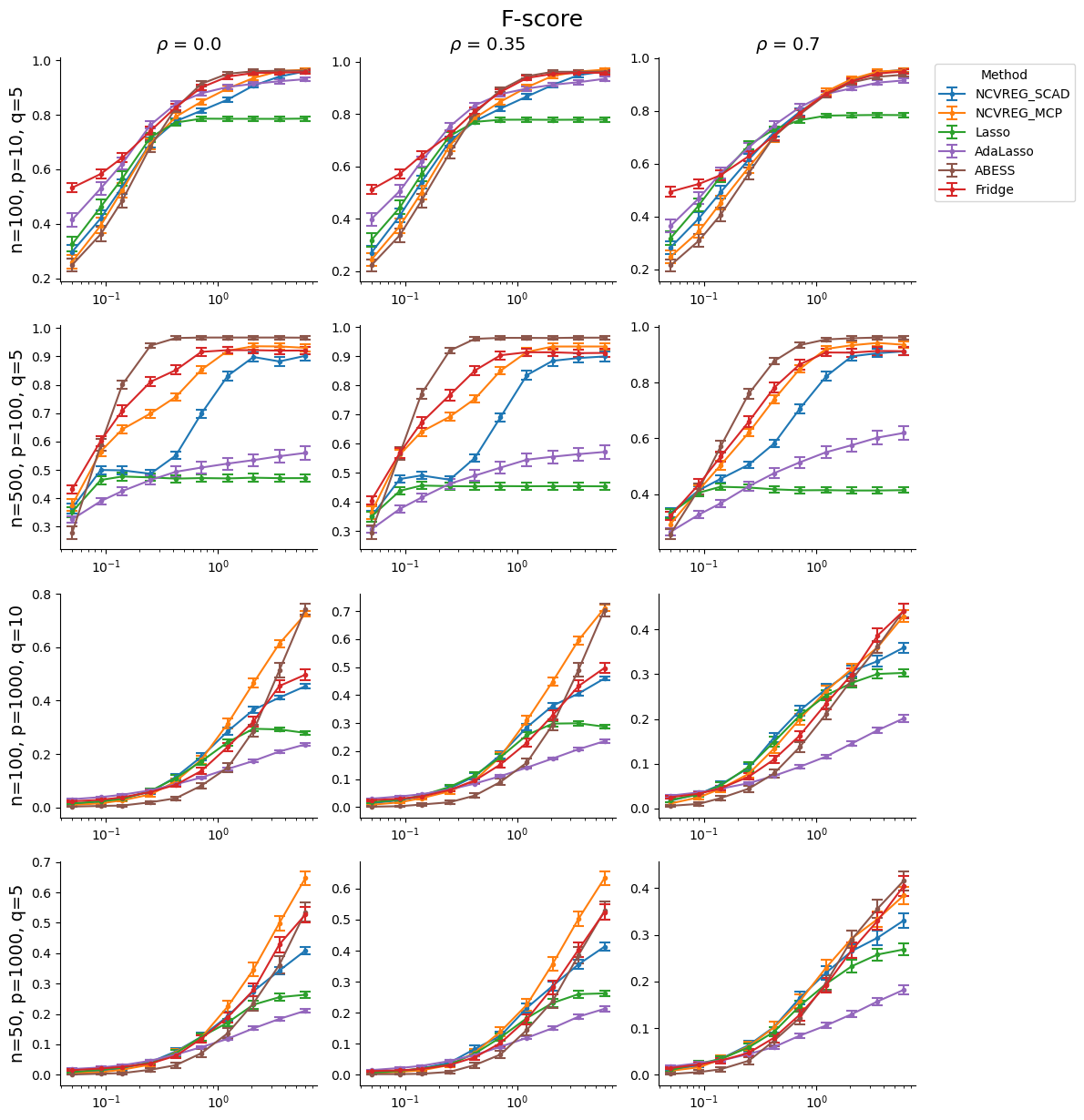}
  \caption[F-score in the setting with step $=2$.]{
    F-score (ordinate) versus SNR (abscissa) for Step $=2$. The ten SNR values range from $0.05$ to $6$ and are equally spaced on a $\log_{10}$ scale. Rows correspond to the indicated $(n,p,q)$ settings, and columns to predictor correlations $\rho=0$, $0.35$, and $0.7$. The ordinate is linear, with axis limits varying across panels. Points show means over 500 independent Monte Carlo replications; error bars indicate $\pm 2$ standard errors.
  }
  \label{fig:step2_fscore}
\end{figure}

\begin{figure}[!htp]
\centering
  \includegraphics[width=\textwidth]{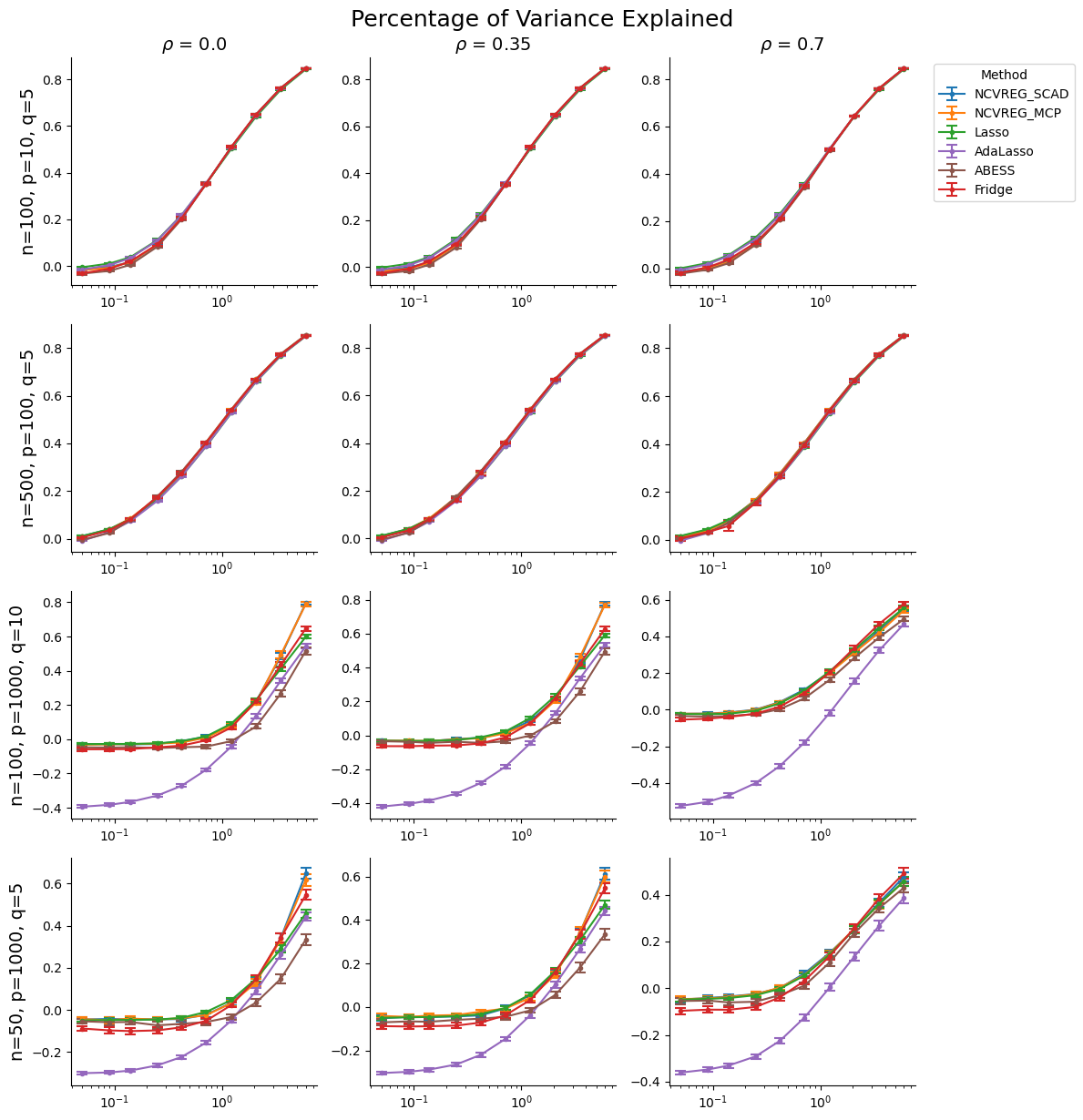}
  \caption[PVE in the setting with step $=2$.]{
    Proportion of variance explained (PVE) (ordinate) versus SNR (abscissa) for Step $=2$. The ten SNR values range from $0.05$ to $6$ and are equally spaced on a $\log_{10}$ scale. Rows correspond to the indicated $(n,p,q)$ settings, and columns to predictor correlations $\rho=0$, $0.35$, and $0.7$. The ordinate is linear, with axis limits varying across panels. Points show means over 500 independent Monte Carlo replications; error bars indicate $\pm 2$ standard errors.
  }
  \label{fig:step2_pve}
\end{figure}

\begin{figure}[!htp]
\centering
  \includegraphics[width=\textwidth]{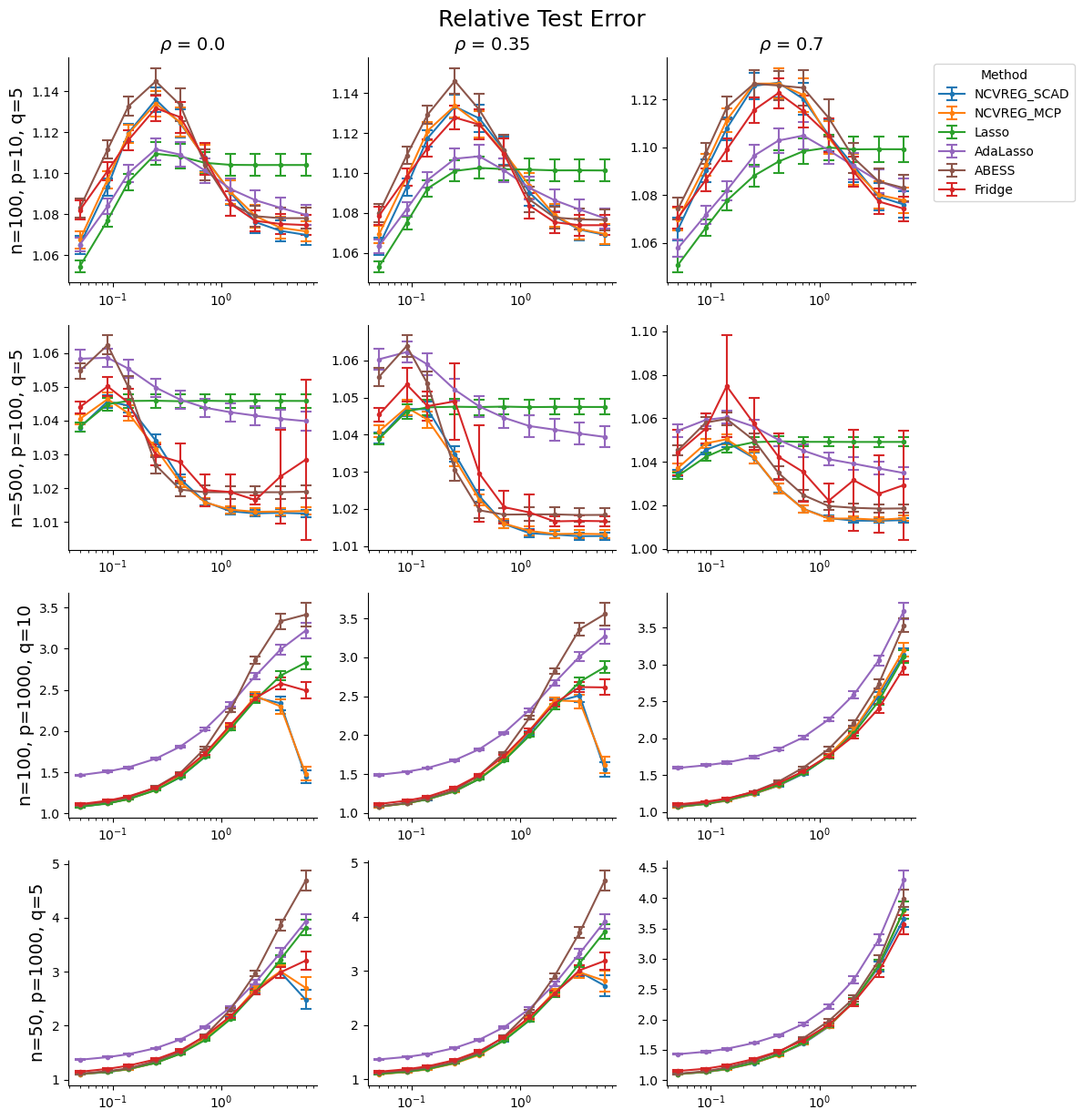}
  \caption[RTE in the setting with step $=2$.]{
    Relative test error (RTE) (ordinate) versus SNR (abscissa) for Step $=2$. The ten SNR values range from $0.05$ to $6$ and are equally spaced on a $\log_{10}$ scale. Rows correspond to the indicated $(n,p,q)$ settings, and columns to predictor correlations $\rho=0$, $0.35$, and $0.7$. The ordinate is linear, with axis limits varying across panels. Points show means over 500 independent Monte Carlo replications; error bars indicate $\pm 2$ standard errors.
  }
  \label{fig:step2_rte}
\end{figure}

\begin{figure}[!htp]
\centering
  \includegraphics[width=\textwidth]{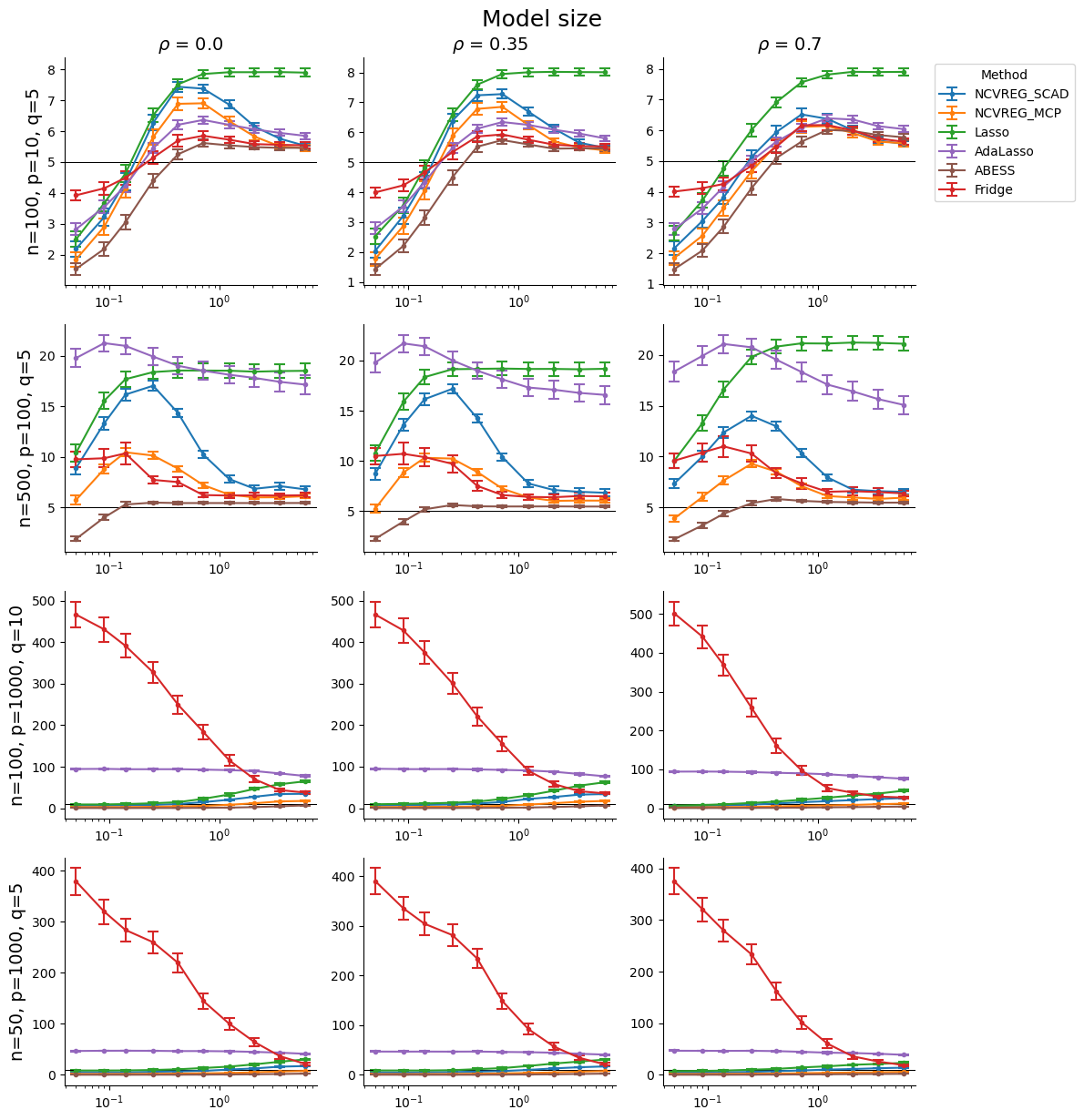}
  \caption[Model size in the setting with step $=2$.]{
    Selected model size (ordinate) versus SNR (abscissa) for Step $=2$. The ten SNR values range from $0.05$ to $6$ and are equally spaced on a $\log_{10}$ scale. Rows correspond to the indicated $(n,p,q)$ settings, and columns to predictor correlations $\rho=0$, $0.35$, and $0.7$. The ordinate is linear, with axis limits varying across panels. Points show means over 500 independent Monte Carlo replications; error bars indicate $\pm 2$ standard errors. The solid black line indicates the true model size $q$.
  }
  \label{fig:step2_modelsize}
\end{figure}

Figures \ref{fig:step1_fscore}--\ref{fig:step2_modelsize} complement the representative simulation results in Section \ref{sec:sim} by reporting all combinations of sample size, predictor dimension, predictor correlation, and spacing between active predictors considered in the study.

\end{document}